\documentclass[iop,numberedappendix]{emulateapj}
\usepackage{graphicx,amsmath,url, color}
\usepackage{times}
\usepackage{bm}
\usepackage{hyperref}
\usepackage{subfigure}
\usepackage{cancel}

\usepackage{color}
\definecolor{brown}{rgb}{0.42,0.24,0.07}
\definecolor{darkgreen}{rgb}{0.0,0.6,0.00}
\definecolor{purple}{rgb}{0.7,0.0,0.7}

\begin{document}

\title{On the axisymmetric stability of stratified and magnetized accretion disks}
\shorttitle{Axisymmetric Stability of Disks}
\author{Gopakumar~Mohandas \& Mart\'{i}n~E.~Pessah}
\shortauthors{Mohandas \& Pessah}
\affiliation{Niels Bohr International Academy, Niels Bohr Institute, Blegdamsvej 17, DK-2100, Copenhagen~\O, Denmark}
\email{gopakumar@nbi.ku.dk, mpessah@nbi.ku.dk}

\begin{abstract}
We conduct a comprehensive axisymmetric, local linear mode analysis of a stratified, differentially rotating disk permeated by a toroidal magnetic field which could provide significant pressure support.  In the adiabatic limit, we derive a new stability criteria that differs from the one obtained for weak magnetic fields with a poloidal component and reduces continuously to the hydrodynamic Solberg-H{\o}iland criteria. Three fundamental unstable modes are found in the locally isothermal limit. They comprise of overstable: {\it i)} acoustic oscillations, {\it ii)}  radial epicyclic (acoustic-inertial) oscillations and {\it iii)}  vertical epicyclic (or vertical shear) oscillations. All three modes are present for finite ranges of cooling times but they are each quickly quenched past respective cut-off times. The acoustic and acoustic-inertial overstable modes are driven by the background temperature gradient. When vertical structure is excluded, we find that the radial epicyclic modes appear as a nearly degenerate pair. One of these is the aforementioned acoustic-inertial mode and the other has been previously identified in a slightly different guise as the convective overstability. Inclusion of vertical structure leads to the development of overstable oscillations destabilized by vertical shear but also has the effect of suppressing the radial epicyclic modes. Although our study does not explicitly account for non-ideal effects, we argue that it may still shed light into the dynamics of protoplanetary disk regions where a strong toroidal field generates as a result of the Hall-shear instability.
\end{abstract}


\section{Introduction}
\label{sec_intro}

Hydrodynamic and magnetohydrodynamic instabilities in accretion disks
play a fundamental role in governing their evolution. 
From the generation of a turbulent viscosity enabling angular momentum
transfer to the formation of vortices and substructure such as spiral arms;
fluid instabilities are thought to be the main drivers
of disk dynamics.
It is therefore vital
to identify the sources and determine the conditions under which
various disk instabilities are present.

For disks threaded by magnetic fields, the magneto-rotational instability (MRI)  \citep{bh1, bh4} 
has long been the front-runner in explaining the mechanism of angular 
momentum transport. 
Without relying on shear, 
certain magnetic field configurations can also utilize
suitable entropy gradients to destabilize parts of the disk. The magnetic
interchange \citep{new61, dja79} and the Parker \citep{enp66, shu74, ft94, ft95}
instabilities are two such modes that could play important roles especially in 
disk atmospheres. 

In protoplanetary disks, where ionization levels are
generally considered to be too low to allow the MRI to operate efficiently within vast swathes of the disk, 
alternative sources of disk turbulence have been looked into
 to explain observed accretion rates \citep{arm11, tbfgklw14}.
Turbulence due to hydrodynamic convection \citep{ms58} is still a contentious
candidate \citep{lp80, rpl88, rg92, lpk93, lo10} when it comes to the
question of facilitating angular momentum transport. 
The vertical shear instability (VSI;
\citealt{vuab98, vu03, ngu13, bl15, cmmp14, ly15}) has received 
much attention lately as a means to generate
turbulence in those parts of the disk that are thought to be
magnetically inactive. Originally studied in the context of rotating stars 
\citep{gs67, fr68}, recent work by \citet{vuab98, vu03} has 
bestowed upon this instability renewed relevance.  
The VSI operates by feeding off of vertical gradients in the angular frequency ensuing
from the baroclinic nature of a disk with radial thermal structure.

Recent observations of protoplanetary disks have revealed
complex internal structures within the disk body \citep{fukaetal13, picc14, aqsmgwd14}. Spiral arms
and vortices are now expected to be common features
of disks around young forming stars  \citep{mutoetal12, ll13, ccmpv14}. 
The sub-critical baroclinic
instability \citep{lp10}, the Rossby wave instability \citep{llcn99, lflc00}, the convective
overstability \citep{kh14, lyra14},  are among a few hydrodynamic
instabilities that have been studied in this regard. The sub-critical baroclinic instability
is a non-linear instability and requires finite amplitude perturbations to generate vortices.
Recent work by \citet{kh14} and \citet{lyra14} have found that 
if the vertical structure of the disk is ignored, a linear convective 
overstability arises and it reaches maximum growth when
the thermal relaxation takes place on a dynamical timescale.
This convective  overstability is thought to be related to the 
subcritical baroclinic instability studied by \citet{lp10}, and has 
been the focus of recent attention because of its potential  
implications for vortex formation in hydrodynamic disks.

There is little doubt that the prevalent physical conditions governing the
dynamical and thermo-dynamical properties of astrophysical disks offer
a number of sources of free energy to drive instabilities.  
Some of these sources become accessible when the
disk is magnetized and sufficiently ionized, e.g., the MRI, while
others work more efficiently when the gas can thermally relax
sufficiently fast, e.g., the VSI.  Previous studies on the
stability properties of disks have usually addressed different aspects
of this problem by focusing on either the hydrodynamic or weakly
magnetized limit and usually working in either the isothermal or
adiabatic regime.

The central aim of this paper is to perform a comprehensive local
stability analysis of a stratified, differentially rotating disk
permeated by a purely toroidal magnetic field, which could be strong
enough to provide significant pressure support.  We consider the
full range of thermal relaxation times, and relax some assumptions
that have been previously invoked in similar linear mode analysis.
This allows us to relate our results to a number of studies and address 
some subtle and outstanding issues.

For adiabatically evolving systems, the Solberg-H{\o}iland (hereafter
SH) criteria \citep{tass} determines the stability to axisymmetric
disturbances in stratified and differentially rotating systems in the
absence of magnetic fields. \citet{bal95} (hereafter BH95)
investigated the condition for stability when weak magnetic fields are
included and derived a modified version of the SH stability criteria
where the gradients in the angular momentum are replaced by angular
velocity gradients.  The BH95 criteria is not directly applicable in
the case of magnetic field configurations that add to pressure support and it contains a singular limit
wherein the criteria for stability does not continuously reduce to the
SH criteria in the limit of vanishingly weak magnetic field.  In this
paper, we derive the criteria that govern the stability of adiabatic
axisymmetric disturbances in the presence of purely toroidal magnetic
fields and explain why in the absence of a poloidal field component
these criteria, that differ from BH95, match smoothly the SH criteria
in the limit of weak fields.

Thin, adiabatic, hydrodynamic disks are quite resilient to local, 
axisymmetric perturbations. Nevertheless, the wide range of
physical conditions in astrophysical disks can host thermodynamic 
processes characterized by a wide range of timescales, 
which can be effectively modeled via finite thermal relaxation.
There are a number of instabilities that can operate
when the thermal relaxation timescale is shorter than, 
or of the order of, the dynamical timescale. The VSI and the 
convective overstability are examples of two such instabilities
that are known to operate within a finite range of thermal
relaxation times. 

Due to the complimentary nature of the approximations involved
in previous studies, it is still unclear whether instabilities like the VSI and the 
convective overstability could appear together and
under what circumstances one or the other dominates. 
In this paper, we present a coherent framework to investigate the 
stability of differentially rotating, stratified and magnetized accretion disks.
In doing so, we are able to identify and characterize the nature 
of the modes that prevail within accessible regions of parameter space
as well as ascertain their relative predominance. We uncover certain
unstable modes, some of which appears to have been previously overlooked.

The plan of the paper is as follows.  In Section \ref{sec_basiceqns},
we state the basic equations of motion and describe an equilibrium
disk model that we shall use for our calculations. In Section
\ref{sec_dispersionrelation}, we derive the general dispersion
relation. In Section \ref{sec_critstability}, we derive the stability
criteria in the adiabatic limit and briefly discuss prospects of
instability based on the given disk model. In Section
\ref{sec_finitecooling}, we solve the dispersion relation for finite 
thermal relaxation times  and derive analytical approximations for the
unstable modes.  In Section \ref{sec_PPD}, we argue about the 
potential relevance of our finding for protoplanetary disks.
In Section \ref{sec_discussion}, we summarize our 
results and discuss them in the context of previous work.

\section{Basic Equations and Equilibrium Disk Model}
\label{sec_basiceqns}
 
We consider a magnetized fluid governed by the equations
\begin{subequations}
\label{basiceqns}
\begin{align}
\label{eq_becont}
\frac{\partial \rho}{\partial t} + \boldsymbol{\nabla} \cdot (\rho \boldsymbol{u}) & = 0, \\
\label{eq_bemom}
\frac{\partial \boldsymbol{u}}{\partial t} + \boldsymbol{u} \cdot \boldsymbol{\nabla} \boldsymbol{u} &= -\frac{1}{\rho} \boldsymbol{\nabla} \left( P + \frac{B^2}{8 \pi} \right) - \boldsymbol{\nabla} \Phi + \boldsymbol{B} \cdot \boldsymbol{\nabla} \boldsymbol{B}, \\ 
\label{eq_beind}
\frac{\partial \boldsymbol{B}}{\partial t} & =  \boldsymbol{\nabla} \times (\boldsymbol{u} \times \boldsymbol{B}), \\
\label{eq_beent}
\frac{\partial P}{\partial t} + \boldsymbol{u} \cdot \boldsymbol{\nabla}  P & = -\gamma P \boldsymbol{\nabla} \cdot \boldsymbol{u} - \Lambda,
\end{align}
where $\rho$ is the density, $\boldsymbol{u}$ is the velocity, $P$ is the pressure, $\boldsymbol{B}$ is the magnetic field, $\Phi$ is the gravitational potential of the central object and $\gamma$ is the adiabatic index. Thermal relaxation in the fluid is modeled by the term $\Lambda$, defined here as
\begin{align}
\Lambda \equiv \frac{P}{T}\frac{T-T_0}{t_c} \,,
\end{align} 
\end{subequations}
where $T_0$ is some reference equilibrium temperature and $t_c$ is the thermal relaxation time. 

We adopt a cylindrical coordinate system $(R, \phi, z)$ throughout the analysis. All the fluid variables are axisymmetric and in general a function of both the radial and vertical coordinates. We consider the background velocity and magnetic field to be purely toroidal
\begin{align}
\boldsymbol{u}  = R \Omega(R,z) \boldsymbol{\hat{\phi}} \,, \\
\boldsymbol{B} = B_0(R, z) \boldsymbol{\hat{\phi}} \,,
\end{align}
where $\Omega$ is the angular frequency.
Additionally, we specify the equation for the
vorticity $\boldsymbol{\omega} = \boldsymbol{\nabla} \times \boldsymbol{u} $,
which is obtained by taking the curl of Equation (\ref{eq_bemom})
\begin{align}
\label{eq_bevor}
\frac{\partial \boldsymbol{\omega}}{\partial t} &= \boldsymbol{\nabla} \times (\boldsymbol{u} \times \boldsymbol{\omega}) + \frac{ \boldsymbol{\nabla} P_{\rm T} \times \boldsymbol{\nabla} \rho}{\rho^2} + \boldsymbol{\nabla} \times \frac{B_0^2}{4 \pi \rho_0 R} \boldsymbol{\hat{R}},
\end{align}

Equilibrium solutions that describe a magnetized, differentially rotating disk can be obtained by solving the set of Equations (\ref{basiceqns}) in the steady state. It is not always a trivial exercise to derive such solutions especially when magnetic fields are to be included. For our purpose here, a useful disk model can be obtained by assuming that the basic fluid variables have the self-similar 
power law form in radius at the disk mid-plane
\begin{align}
\label{rhorad}
\rho(R,0) &= \rho_{00} \left( \frac{R}{R_0} \right)^p \,,  \\
\label{brad}
B(R,0) &= B_{00} \left( \frac{R}{R_0} \right)^{p/2} \,,  \\
\label{trad}
T(R) &= T_{00} \left( \frac{R}{R_0} \right)^q \,,
\end{align}
where $R_0$ is a fiducial radius.
We furthermore assume that the Alfv\'{e}n speed, $c_{\rm A}$, is constant.
We also assume an ideal gas equation of state
\begin{align}
P = \frac{\mathcal{R}_{\rm g}}{\mu_{\rm g} } \rho T \, ,
\end{align}
where $\mathcal{R}_{\rm g}$ is the universal gas constant and $\mu_{\rm g}$ 
the mean molecular weight of the gas.
Assuming a steady state, substituting Equations (\ref{rhorad})-(\ref{trad}) 
in the momentum Equation (\ref{eq_bemom}) and integrating, we obtain
\begin{align}
\label{dm_rho}
\rho(R,z) &= \rho_{00} \left( \frac{R}{R_0} \right)^p \exp \left[ \frac{G M}{(c_{\rm i}^2 + c_{\rm A}^2)} \left( \frac{1}{\sqrt{R^2 + z^2}} - \frac{1}{R} \right) \right],   \\
\label{dm_b}
B(R,z) &= c_{\rm A} \sqrt{ 4 \pi \rho (R,z)}  \,.
\end{align} 
The radial temperature gradient leads to an angular frequency profile that is not constant on cylinders and consequently has the form 
\begin{eqnarray}
\label{dm_ang}
\frac{\Omega^2 (R,z)}{\Omega_{\rm K}^2(R)} = 1 &+& (p + q)\frac{c_{\rm i}^2}{v_{\rm K}^2} + \left(1+\frac{p}{2} \right)\frac{c_{\rm A}^2}{v_{\rm K}^2} \nonumber \\
&+& \frac{q c_{\rm i}^2}{(c_{\rm i}^2 + c_{\rm A}^2)} \left( 1 - \frac{R}{\sqrt{R^2 + z^2}} \right) \,,
\end{eqnarray}
where $v_{\rm K} = R \Omega_{\rm K}$ is the Keplerian angular velocity and $c_{\rm i} = \sqrt{P/\rho}$, the isothermal sound speed. 
In the absence of magnetic fields, the equilibrium solutions (\ref{dm_rho}) and (\ref{dm_ang}) reduce to solutions used previously in the study of protoplanetary disks (see for example \citealt{ngu13}). The magnetized solutions (\ref{dm_rho})- (\ref{dm_ang}) have been used previously by \citet{pb13a, pb13b} for global studies of the MRI. 

\begin{table}
\caption[]{List of symbols used in this paper.}
\label{table:symbols}
\begin{center}
\begin{tabular}{l l l}\hline
Notation & Definition & Description \\\hline
\vspace{1mm}
$k_R$ &    & Radial wavenumber \\
\vspace{1mm}
$k_z$ &  & Vertical wavenumber \\
\vspace{1mm}
$k$ &$=\sqrt{k_R^2 + k_z^2}$& Total wavenumber  \\
\vspace{1mm}
$\mu$ &$=k_z/k $&  \\
\vspace{1mm}
$\Omega_{\rm K}$ & $\equiv \sqrt{G M/R^3}$ & Keplerian angular frequency \\
\vspace{1mm}
$\omega$ &$ = \nu + i \sigma$ & Complex eigenvalue \\
\vspace{1mm}
$\nu$  &$=Re[\omega]$ & Oscillation frequency of mode \\
\vspace{1mm}
$\sigma$ &$=Im[\omega]$ & Growth rate of mode  \\
\vspace{1mm}
$t_c$ & & Cooling/Thermal relaxation time \\
\vspace{1mm}
$\varsigma$ &$=1/(\gamma t_{\rm c})$ & \\
\vspace{1mm}
$\tau$ &$=t_{\rm c} \Omega_{\rm K}$& Dimensionless cooling time \\
\vspace{1mm}
$c_{\rm i}$ &$\equiv \sqrt{P_0/\rho_0}$ & Isothermal sound speed \\
\vspace{1mm}
$c_{\rm s}$ &$\equiv \sqrt{\gamma} c_{\rm i}$ & Adiabatic sound speed \\
\vspace{1mm}
$c_{\rm A}$ &$\equiv B_0/\sqrt{4\pi \rho_0}$ & Alfv\'{e}n speed  \\
\vspace{1mm}
$H$  & $\equiv c_{\rm i}/\Omega_{\rm K}$ & Density scale height \\
\vspace{1mm}
$h$ & $\equiv H/R $& Disk aspect ratio \\
\vspace{1mm}
$\beta$ & $\equiv 2 c_{\rm i}^2/c_{\rm A}^2 $ &Mid-plane plasma beta \\
\vspace{1mm}
\vspace{1mm}
$P_{\rm T}$ &$\equiv P_0 + B_0^2/8 \pi$& Total pressure (gas + magnetic) \\
\vspace{1mm}
$\boldsymbol{\omega}$ & $\equiv \boldsymbol{\nabla} \times \boldsymbol{u} $ & Vorticity  \\
\vspace{1mm}
\vspace{1mm}
$\kappa_R$ &$\equiv R^{-3} \partial_R (R^4 \Omega_0^2) $& Radial Epicyclic Frequency \\
\vspace{1mm}
$\kappa_z$ &$\equiv R^{-3} \partial_z (R^4 \Omega_0^2) $& Vertical Epicyclic Frequency \\
\vspace{1mm}
\vspace{1mm}
$s$ &$\equiv \ln \left( P_0/\rho_0^{\gamma} \right)$ & Specific entropy \\
\vspace{1mm}
\vspace{1mm}
$m$ &$\equiv \ln \left( B_0/ \rho_0 \right)$  & \\
\vspace{1mm}
\vspace{1mm}
$N_R^2$ &$\equiv - (\partial_R P_T \, \partial_R s)/\gamma \rho_0 $ & Radial Buoyancy Frequency \\
\vspace{1mm}
$N_z^2$ &$\equiv - (\partial_z P_T \, \partial_z s)/\gamma \rho_0 $ & Vertical Buoyancy Frequency \\
\vspace{1mm}
$L_R^2$ &$\equiv - 2(\partial_R P_T \, \partial_R m)/\gamma \beta \rho_0$ & \\
\vspace{1mm}
$L_z^2$ &$\equiv - 2(\partial_z P_T \, \partial_z m)/\gamma \beta \rho_0$ & \\
\vspace{1mm}
$M_R^2$ &$\equiv N_R^2 + L_R^2$ & Radial Magneto-Buoyancy Freq.\\
\vspace{1mm}
$M_z^2$ &$\equiv N_z^2 + L_z^2$ & Vertical Magneto-Buoyancy Freq. \\
\hline
\end{tabular}
\end{center}
\end{table}

\section{Dispersion Relation}
\label{sec_dispersionrelation}

Consider $f$ to be any of the aforementioned fluid variables in Equation (\ref{basiceqns}). We decompose $f$ into a sum of the background steady-state and an infinitesimal perturbation such that
\[
f = f_0 + \delta f.
\]
The 0-subscripted variables represent the background state and the $\delta$-prefixed variables represent the perturbations  above and also in what follows. We linearize Equations (\ref{basiceqns}) by retaining only terms up to first order in the perturbations. We carry out the local linear mode analysis in the WKB approximation where $k_R R \gg 1$ and $k_z z \gg 1$ with the perturbations having the space-time dependence $\exp{( i k_R R + i k_z z - i \omega t )}$. The linearized equations can then be expressed as\footnote{Note that the linearized momentum equation does not contain the term associated with magnetic tension induced by the curvature of
the background toroidal magnetic field.
This is usually ignored in local studies, including the shearing box, 
but could be relevant for very strong toroidal magnetic fields.
We comment on this in more detail below.}
\begin{subequations}
\label{lineqns}
\begin{align}
\label{eq_linur}
i \omega \delta u_R  & = i k_R \frac{\delta p}{\rho_0} + i k_R c_{\rm A}^2 \frac{\delta b_{\phi}}{B_0} - 2 \Omega_0 \delta u_{\phi} - \frac{\delta \rho}{\rho_0} \left( \frac{1}{\rho_0}\frac{\partial P_{\rm T}}{\partial R} \right),
\end{align}
\begin{align}
\label{eq_linuy}
i \omega \delta u_{\phi} &=  \frac{\kappa_R^2}{2 \Omega_0}\delta u_R + \kappa_z^2 \delta u_z,  
\end{align}
\begin{align}
\label{eq_linuz}
i \omega \delta u_z &= i k_z \frac{\delta p}{\rho_0}  + i k_z c_{\rm A}^2  \frac{b_{\phi}}{B_0} - \frac{\delta \rho}{\rho_0} \left( \frac{1}{\rho_0}\frac{\partial P_{\rm T}}{\partial z} \right), 
\end{align}
\begin{align}
\label{eq_linby}
i \omega  \frac{\delta b_{\phi}}{B_0} &= i k_R \delta u_R + i k_z \delta u_z + \delta u_R \frac{\partial \ln B_0}{\partial R} + \delta u_z \frac{\partial \ln B_0}{\partial z},  
\end{align}
\begin{align}
\label{eq_lincont}
i \omega \frac{\delta \rho}{\rho_0} &=  i k_R \delta u_R + i k_z \delta u_z + \delta u_R \frac{\partial \ln \rho_0}{\partial R} + \delta u_z \frac{\partial \ln \rho_0}{\partial z},   
\end{align}
\begin{align}
\label{eq_linent}
i \omega \frac{\delta p}{\rho_0} &= i \omega \frac{\delta \rho}{\rho_0} c_{\rm s}^2 + \left( \frac{1}{\rho_0}\frac{\partial P_0}{\partial R} \right) \delta u_R - \left( \frac{1}{\rho_0}\frac{\partial \rho_0}
{\partial R} \right) c_{\rm s}^2 \delta u_R 
\nonumber \\
& + \left( \frac{1}{\rho_0}\frac{\partial P_0}{\partial z} \right) \delta u_z - \left( \frac{1}{\rho_0}\frac{\partial \rho_0}{\partial z} \right)c_{\rm s}^2 \delta u_z \nonumber \\
& - \frac{1}{t_{\rm c}}\left( \frac{\delta p}{\rho_0} - c_{\rm i}^2\frac{\delta \rho}{\rho_0} \right)   .
\end{align}
\end{subequations}

The set of Equations (\ref{lineqns}) lead to the dispersion relation
\begin{align}
\label{fulldisprel}
 \omega^5 + \frac{i}{t_{\rm c}}  \omega^4 - F_3 \omega^3 - \frac{i}{t_{\rm c}} F_4 \omega^2 + F_5 \omega + \frac{i}{t_{\rm c}} F_6 = 0,
\end{align}
with coefficients 
\begin{subequations}
\label{fullcoefficients}
\begin{align}
\label{c_f3}
F_3 &=  (k_R^2 + k_z^2)\left( c_{\rm s}^2 + c_{\rm A}^2   \right) + \kappa_R^2  \nonumber \\
&
+  \frac{1}{\rho_0^2} \left(\frac{\partial P_{\rm T}}{\partial z}\frac{\partial \rho_0}{\partial z} 
+  \frac{\partial P_{\rm T}}{\partial R} \frac{\partial \rho_0}{\partial R} \right) \,,  \\
\label{c_f4}
F_4 &=   (k_R^2 + k_z^2)\left( c_{\rm i}^2 + c_{\rm A}^2   \right) + \kappa_R^2  
 + i k_R \frac{\partial c_{\rm i}^2}{\partial R}   + i k_z \frac{\partial c_{\rm i}^2}{\partial z}   \nonumber \\
&+  \frac{1}{\rho_0^2} \left(\frac{\partial P_{\rm T}}{\partial z}\frac{\partial \rho_0}{\partial z} 
+  \frac{\partial P_{\rm T}}{\partial R} \frac{\partial \rho_0}{\partial R} \right) \,,  \\
 \label{c_f5}
F_5 &= \kappa_{R}^2 \left[ k_z^2\left( c_{\rm s}^2 + c_{\rm A}^2   \right)  + \frac{1}{\rho_0^2}\frac{\partial P_{\rm T}}{\partial z}\frac{\partial \rho_0}{\partial z}  \right]  \nonumber \\
& - \kappa_{z}^2 \left[ k_R k_z \left( c_{\rm s}^2 + c_{\rm A}^2   \right)  +  \frac{1}{\rho_0^2}\frac{\partial P_{\rm T}}{\partial z}\frac{\partial \rho_0}{\partial R}   
   \right]  \nonumber \\
& + \mathcal{K} \left\{ \left[
-\frac{k_z c_{\rm s}^2}{\gamma} \frac{\partial s}{\partial R}  
- k_z c_{\rm A}^2 \frac{\partial m}{\partial R} \right. + \frac{k_R c_{\rm s}^2}{\gamma} \frac{\partial s}{\partial z} 
+ k_R c_{\rm A}^2 \frac{\partial m}{\partial z}         \right]  \nonumber \\ 
& \left. + \frac{i}{\rho_0^2} \left[ \frac{\partial P_{\rm T}}{\partial R}\frac{\partial \rho_0}{\partial z} - 
 \frac{\partial P_{\rm T}}{\partial z}\frac{\partial \rho_0}{\partial R} + R \frac{\partial \Omega_0^2}{\partial z} \right] \right\} ,  \\
\label{c_f6}
 F_6 & = \kappa_{R}^2 \left[ k_z^2\left( c_{\rm i}^2 + c_{\rm A}^2   \right)  +  i k_z\frac{\partial c_{\rm i}^2}{\partial z} 
 + \frac{1}{\rho_0^2}\frac{\partial P_{\rm T}}{\partial z}\frac{\partial \rho_0}{\partial z}  \right]  \nonumber \\
& - \kappa_{z}^2 \left[ k_R k_z \left( c_{\rm i}^2 + c_{\rm A}^2   \right)  +  \frac{1}{\rho_0^2}\frac{\partial P_{\rm T}}{\partial z}\frac{\partial \rho_0}{\partial R}   \right. \nonumber \\
 & \left.
+  \frac{i k_R}{\rho_0}\frac{\partial P_{\rm T}}{\partial z}  
 -  \frac{i k_z c_{\rm i}^2}{\rho_0} \frac{\partial \rho_0}{\partial R} -   \frac{i k_z c_{\rm A}^2}{B_0} \frac{\partial B_0}{\partial R}    \right]  \nonumber \\
& + \mathcal{K}  \left\{ \left[
k_R c_{\rm A}^2 \frac{\partial m}{\partial z}  - k_z c_{\rm A}^2 \frac{\partial m}{\partial R} \right.    \right]  \nonumber \\  
& \left. + i c_{\rm A}^2 \left[ \frac{\partial \ln B_0}{\partial R} \frac{\partial \ln \rho_0}{\partial z} - 
\frac{\partial \ln B_0}{\partial z} \frac{\partial \ln \rho_0}{\partial R} \right] \right\} .
\end{align}
\end{subequations}
All the major symbols used in these equations  and those that will be used later in the paper have been collated and described in Table \ref{table:symbols}. For ease of notation, we have also defined the variable $\mathcal{K}$ above to represent the collection of terms
\begin{align}
\mathcal{K} = \frac{k_z}{\rho_0} \frac{\partial P_T}{\partial R} - \frac{k_R}{\rho_0} \frac{\partial P_T}{\partial z} \,.
\end{align}

Analytical progress is hampered when considering the dispersion relation (\ref{fulldisprel}) in full generality. The coefficients in  this equation are not all real valued, therefore, extracting a general stability criteria is a complicated exercise. Nonetheless, some headway can be made if we consider certain limits.

\subsection{Approximations and Constraints}

Prior to determining the conditions of stability, we first comment
on the virtues and limitations associated with the assumptions that 
we make and how they impact our analysis.

The imaginary part of the coefficient $F_5$ adds up to the steady 
state vorticity equation if the effect of magnetic tension induced by 
the finite curvature of the background field is neglected. 
Therefore, the dispersion relation that we have derived applies
provided that the background disk model satisfies
\begin{align}
\label{eq_linvort}
\frac{1}{\rho_0^2}\frac{\partial P_{\rm T}}{\partial R}\frac{\partial \rho_0}{\partial z} - \frac{1}{\rho_0^2} \frac{\partial P_{\rm T}}{\partial z}\frac{\partial \rho_0}{\partial R} + R \frac{\partial \Omega_0^2}{\partial z} = 0 \,.
\end{align}
With regard to the vorticity equation, \citet{ras02} pointed out that gas disks with magnetic fields do not in general satisfy Equation (\ref{eq_bevor}) in the steady state. This is because magnetic forces are not conservative and cannot, therefore, be derived from a scalar potential. 
However, this constraint can be overcome for a purely toroidal field if the Alfv\'{e}n speed is a constant\footnote{The vorticity equation is trivially satisfied if the field itself is spatially uniform and such a field configuration with a consistent hydrostatic model would also constitute an equilibrium solution.}. 
In this case, the magnetic field will only act to provide additional pressure support against the inertial forces and the field strength follows the gas density. As far as the local analysis is concerned, terms associated with magnetic field curvature can be ignored in the WKB approximation provided the Alfv\'{e}n speed does not exceed
the geometric mean of the thermal and Keplerian speeds \citep{pp05}. 

If the background state is such that the Alfv\'{e}n speed is a constant, the set of imaginary terms in $F_6$ which make up the magnetic part of Equation (\ref{eq_linvort}) also add up to zero
\begin{align}
\label{eq_linbvort}
\frac{\partial \ln B_0}{\partial R} \frac{\partial \ln \rho_0}{\partial z} - \frac{\partial \ln B_0}{\partial z} \frac{\partial \ln \rho_0}{\partial R} = 0.
\end{align} 

\section{Stability Criteria in the Adiabatic Limit}
\label{sec_critstability}

In the adiabatic limit, the set of assumptions outlined above lead to a dispersion relation with real 
coefficients, which allows us to derive a criterion for stability straightforwardly. 
The adiabatic limit can be reached by taking $t_c \rightarrow \infty$, and 
the dispersion relation (\ref{fulldisprel}) reduces to
\begin{align}
\label{eq_addisp}
\omega^4 - F_3 \omega^2 + F_5 = 0,
\end{align}
where the coefficients are given by Equations (\ref{c_f3}) and (\ref{c_f5}). 
The necessary and sufficient conditions for stability are that the roots of the dispersion relation (\ref{eq_addisp}) given by
\begin{align}
\omega = \pm \left[ \frac{ F_3 \pm \sqrt{F_3^2 - 4 F_5} }{2}  \right]^{1/2},
\end{align}
are real. This is true if $F_3$ and $F_5$ are both real and also $F_3^2 > 4 F_5 > 0$ is satisfied. For Rayleigh stable flows$F_3^2 > 0$ and in the WKB approximation, $F_3^2 > 4 F_5$. Therefore, the necessary and sufficient condition for stability is that $F_5 > 0$. With a little rearrangement of the coefficient (\ref{c_f5}), we can express this condition on $F_5$ as
\begin{eqnarray}
\label{eq_quadcond}
&& \frac{k_R^2}{k_z^2} M_{\rm z}^2 + M_R^2 + \kappa_0^2\left( 1 + \frac{c_{\rm A}^2}{c_{\rm s}^2} \right) \nonumber \\
 &+& \frac{k_R}{k_z} \left[ \frac{2}{\gamma \rho_0} \frac{\partial P_{\rm T}}{\partial z} \frac{\partial s}{\partial R}  + \frac{2 c_{\rm A}^2}{\rho_0 c_{\rm s}^2 } \frac{\partial P_{\rm T}}{\partial z} \frac{\partial m}{\partial R}  \right] \nonumber \\
&+& \frac{1}{k_z^2 (c_{\rm s}^2 + c_{\rm A}^2) \rho_0^2}\frac{\partial P_{\rm T}}{\partial z}\left( \kappa_0^2 \frac{\partial \rho_0}{\partial z} - R \frac{\partial \Omega_0}{\partial z} \frac{\partial \rho_0}{\partial R} \right) > 0 \,. \,\,\,\,\, 
\end{eqnarray}

In the WKB approximation, the last term on the left hand side of the inequality (\ref{eq_quadcond}) can be dropped leaving behind a simple quadratic expression in $k_R/k_z$. The two conditions that guarantee $F_5 > 0$ are then, {\it i)} the quadratic expression be positive for some value of $k_R/k_z$, and {\it ii)} the quadratic discriminant in (\ref{eq_quadcond}) is negative. These two conditions are tantamount to requiring that 
\begin{subequations}
\label{mshstabcrit}
\begin{align}
\label{eq_mshcrit1}
&M_R^2 + M_z^2 + \kappa_R^2 \left( 1 + \frac{2}{\gamma \beta} \right) > 0,  \\
\label{eq_mshcrit2}
& \frac{\partial P_{\rm T}}{\partial z} \left\{ \left[ \kappa_R^2 \frac{\partial s}{\partial z}  - \kappa_z^2 \frac{\partial s}{\partial R}  \right]  +  \frac{2}{\beta} \left[ \kappa_R^2 \frac{\partial m}{\partial z}   - \kappa_z^2 \frac{\partial m}{\partial R}  \right]  \right\} < 0.
\end{align}
\end{subequations}
Inequalities (\ref{mshstabcrit}) are the generalized version of the Solberg-H{\o}iland criterion with a purely toroidal magnetic field. The above derived stability criteria differs notably from the criteria derived by \citet{bal95} for rotating stratified systems in the presence of a weak magnetic field, which is expressed below to ease the comparison,
\begin{align}
\label{eq_b95crit1}
N_R^2 + N_z^2 + \frac{\partial \Omega^2}{\partial \ln R} > 0 \,,  \\
\label{eq_b95crit2}
\frac{\partial P_0}{\partial z} \left[ \frac{\partial \Omega^2}{\partial R} \frac{\partial s}{\partial z} - \frac{\partial \Omega^2}{\partial z} \frac{\partial s}{\partial R}  \right] < 0 \,.
\end{align}

The differences arise due to the following reasons. The magnetic field considered here provides an appreciable measure of support against the centrifugal forces and gravity while the magnetic field considered in \citet{bal95} is weak in this respect. Furthermore, axisymmetric perturbations to the background toroidal field have the virtue of also being purely toroidal in the ideal limit. As a result, there are no terms containing the radial or vertical shear rate entering the induction equation. This peculiarity is solely responsible for the absence of the singular limit in the criteria (\ref{mshstabcrit}) that is otherwise present in the criteria derived by \citet{bal95}. It is for the same reason that the MRI stability criterion does not emerge out of Equations (\ref{mshstabcrit}). Therefore, as we approach the limit $\beta \rightarrow \infty$, the dispersion relation (\ref{eq_addisp}) reduces to equation (32) of \citet{ras02}, and the stability criteria (\ref{mshstabcrit}) reduce to the more familiar Solberg-H{\o}iland criterion 
\begin{subequations}
\label{shstabcrit}
\begin{align}
\label{eq_shcrit1}
N_R^2 + N_z^2 + \kappa_R^2 > 0 ,  \\
\label{eq_shcrit2}
\frac{\partial P_0}{\partial z} \left[ \kappa_R^2 \frac{\partial s}{\partial z} - \kappa_z^2 \frac{\partial s}{\partial R}  \right] < 0.
\end{align}
\end{subequations}

In the case of a non-rotating system, the new criteria given by Equations (\ref{mshstabcrit}) reduce even further to
\begin{align}
\label{eq_intercrit}
M_R^2 + M_z^2 > 0,
\end{align}
and in the absence of magnetic fields to
\begin{align}
\label{eq_convcrit}
N_R^2 + N_z^2 > 0.
\end{align}
Inequalities (\ref{eq_intercrit}) and (\ref{eq_convcrit}) are the basic form of the interchange \citep{dja79} and the convective \citep{ms58} stability criteria. The interchange instability may be thought of as the magnetized version of the convective instability in which gravity modes are modulated by the magnetic field. These modes are different from the Parker modes as they do not lead to local deformations of the field \citep{kfm98}. In the absence of stratification in Equations (\ref{mshstabcrit}), we recover the Rayleigh criteria for adiabatic perturbations, $\kappa_R^2 > 0$.

Bearing in mind the limitations that were mentioned in the previous section, it is important to emphasize that the stability 
criteria that we have derived in this section is general  in the sense that it is independent of specific attributes of a particular disk model. We proceed below to apply the stability criteria to the specific model
introduced in Section~\ref{sec_basiceqns}.

\subsection{Stability Criteria in Terms of Disk Model}

We examine the possibility of instability in the adiabatic limit for a disk modeled by the equilibrium solutions, Equations (\ref{adm_drdz}) - (\ref{adm_kapz}). An unmagnetized disk can only be unstable to adiabatic perturbations if either one or both of the SH criteria given by Equation (\ref{shstabcrit}) is violated. In the cylindrical disk model of Section \ref{sec_basiceqns}, both $\kappa_R > 0$ and $N_z^2 > 0$. The square of the radial buoyancy frequency is negative if, expressed in terms of the disk model parameters, $p (1 - \gamma) + q < 0$. Standard values of the power law indices are generally of order unity. It is certainly possible to find values within acceptable regimes of parameter space (see next section) where $N_R^2 < 0$. However, since it is likelier that $|\kappa_R^2| \sim \Omega_{\rm K}^2   \gg |N_R^2|$, the first SH condition for stability may still be satisfied. 

According to the second SH stability criterion, the disk model is stable if the following condition 
on the parameters is satisfied
\begin{align}
\label{adhydpiece}
 \frac{(\gamma - 1)}{h^2} > q [p(1-\gamma) + q] \,.
\end{align}
Stability is ensured for the standard values of the parameters usually employed to model protoplanetary disk
if the disk is very thin. With $p$ and $q$ of order unity and with $\gamma \simeq 1.4$ appropriate for protostellar disks, adiabatic destabilization becomes unlikely for very thin disks, $h \ll 1$.

Moreover, for the particular disk model that we consider
\begin{align}
\frac{\partial m}{\partial z} = -\frac{1}{2}\frac{\partial \ln \rho}{\partial z} > 0 ,\\
\frac{\partial m}{\partial R} = -\frac{1}{2}\frac{\partial \ln \rho}{\partial R} > 0,
\end{align}
and thus
\begin{align}
\label{admagpiece}
\frac{2}{\beta} \left[ \kappa_R^2 \frac{\partial m}{\partial z}   - \kappa_z^2 \frac{\partial m}{\partial R}  \right]   > 0.
\end{align}
Equation (\ref{admagpiece}) grows incrementally positive with increasing magnetic field strength. Therefore, even in cases where Equation (\ref{adhydpiece}) were not to be satisfied, a substantially strong field could counteract and ensure stability.

\section{Unstable Modes in the Finite Cooling Regime}
\label{sec_finitecooling}

A realistic disk is expected to possess thermal relaxation times that are neither zero nor infinite 
but fall within a finite range that is set by the prevailing physical conditions. The dispersion relation 
associated with finite, non-zero cooling times is rather involved and finding full analytic solutions is
thus very challenging. In this section we take two parallel paths by solving the dispersion relation 
numerically and working with suitable approximations that allow us to gain insight into specific unstable 
modes. This approach enables us to address a number of subtleties associated with some of the 
approximations that have been previously invoked and to uncover new instabilities that emerge when 
these are relaxed.

The equilibrium solutions presented in Section \ref{sec_basiceqns} 
can be simplified in the thin disk approximation so that, for instance, the normalized density profile has the approximate form
\begin{align}
\label{adm_rho}
\rho_0 &=  \left( \frac{R}{R_0} \right)^p \exp \left[ -\frac{(z-z_0)^2}{2 H^2}\frac{\beta}{(1+\beta)} \right]   .
\end{align}
The vertical gradients in the disk model described in Section \ref{sec_basiceqns} vary linearly with height in the thin disk approximation.
For simplicity, we treat the vertical derivatives as being effectively constant within a small region around some
fiducial height, $z_0$. This treatment is akin to the point analysis approach adopted in \citet{gs67,ngu13} in their
study of the VSI. This approach is valid so long as one stays within the limits of the WKB approximation.
The spatial derivatives that are present in the coefficients of the dispersion relation can then be approximated, in the limit $R -R_0 \ll R_0, z-z_0 \ll R_0$, as follows
\begin{align}
\label{adm_drdz}
R \frac{\partial \ln \rho_0}{\partial z} &=2R \frac{\partial \ln B_0}{\partial z}  = -\frac{n \beta}{h(1+\beta)}  \, ,  \\
\label{adm_drdr}
\frac{\partial \ln \rho_0}{\partial \ln R} &= 2\frac{\partial \ln B_0}{\partial \ln  R} = p \, ,  \\
\label{adm_dptdz}
 \frac{R}{\rho_0}\frac{\partial P_{\rm T}}{\partial z} &= -\frac{n c_{\rm i}^2}{h} \, ,  \\
\label{adm_dptdr}
\frac{R}{\rho_0}\frac{\partial P_{\rm T}}{\partial R} &= \left[ q+ \frac{p \beta}{(1 + \beta)} \right] c_{\rm i}^2 \, 
\end{align}
where $(z-z_0)/H = n$ and $n \geq 0$. We take the radial epicyclic frequency to be equal to the Keplerian 
angular frequency,  $\kappa_R^2 = \Omega_{\rm K}^2$. In the thin disk approximation, the vertical 
epicyclic frequency is approximated by 
\begin{align}
\label{adm_kapz}
\kappa_z^2 \simeq \frac{q n h \beta}{(1 + \beta)} \Omega_{\rm K}^2.
\end{align}

The criteria for neglecting the magnetic tension induced by the curvature of the 
background magnetic field according to \citet{pp05}, may 
be expressed as
\begin{align}
\label{eq_bbound2}
\beta \gg 2 h.
\end{align}
We  consider a disk with an aspect ratio of $h = 0.05$ at some fiducial location, thus, even a plasma beta value of $\beta \sim 10$ obeys the restrictions prescribed by Equation (\ref{eq_bbound2}). The other disk model parameters that we use throughout for all the calculations performed in the rest of this paper are $(p, q, \gamma) = (-2,-1,1.4)$.

\begin{figure*}
\centering  
\includegraphics[width=0.495\textwidth]{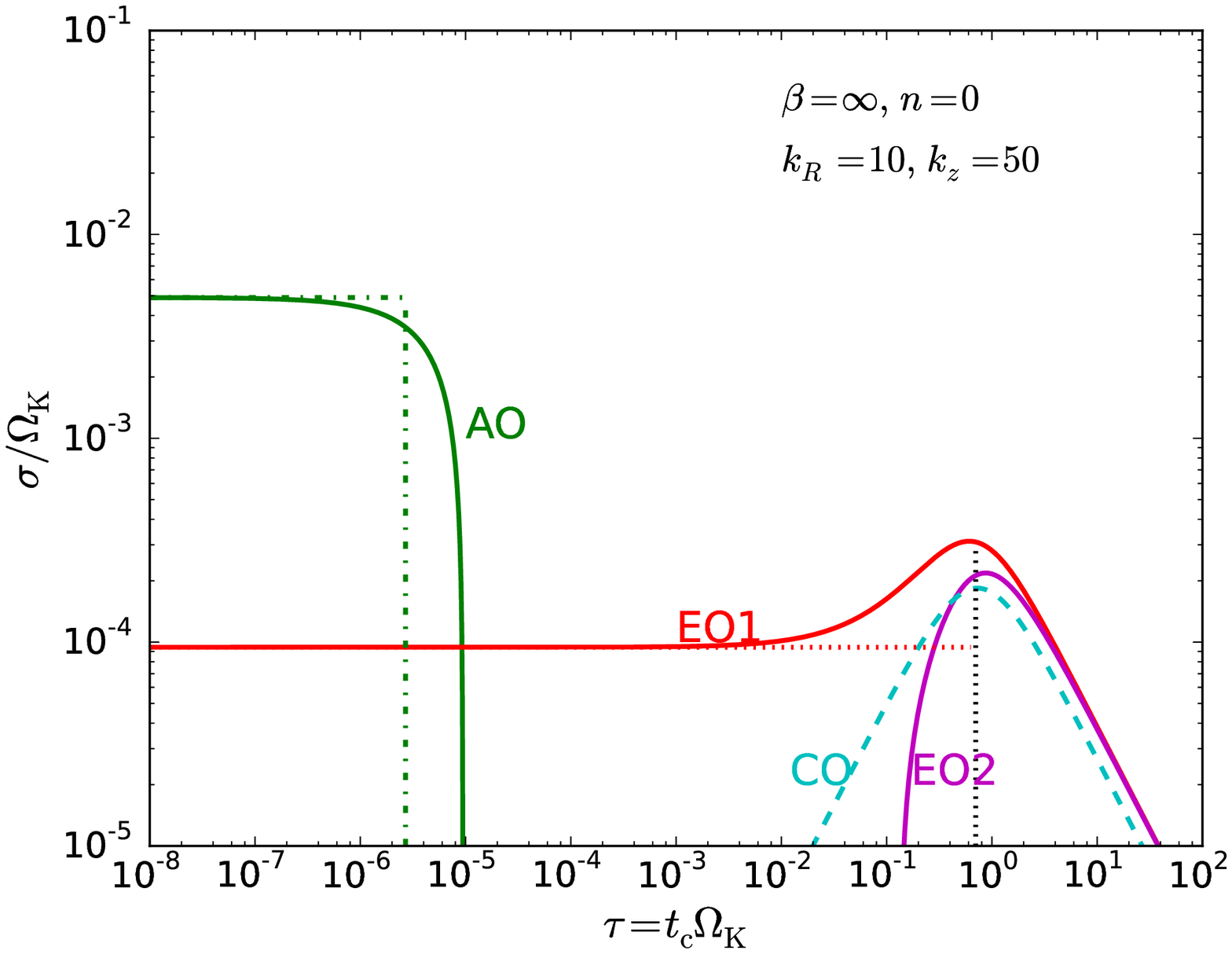}
\includegraphics[width=0.495\textwidth]{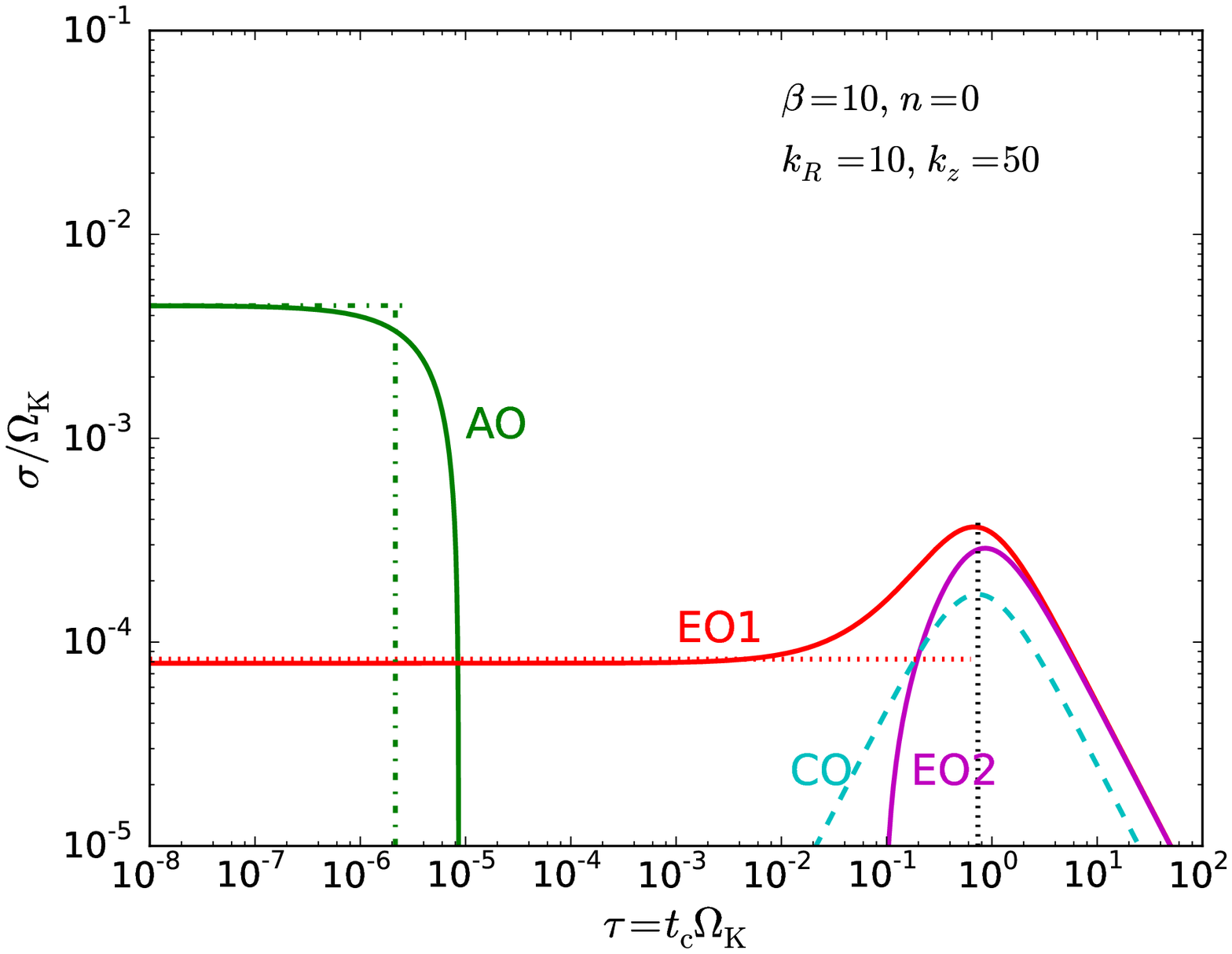}   
\caption{Growth rate of the unstable mode versus thermal relaxation time. These solutions have been obtained without including vertical structure. The disk model parameters used are $(p, q, h, \gamma) = (-2, -1, 0.05, 1.4)$. The unstable growth rate $\sigma$ and the cooling time $\tau$ are non-dimensionalized in terms of $\Omega_{\rm K}$. The abbreviations used are: AO - Acoustic Overstability, EO1 - first radial epicyclic Overstability and EO2 - second radial epicyclic overstability \\
} 
\label{n0gvt}
\end{figure*}

Using Equations (\ref{adm_drdz})-(\ref{adm_kapz}) in the coefficients of Equation (\ref{fulldisprel}), we numerically compute 
the roots of the dispersion relation. 
Figures \ref{n0gvt}, \ref{n2gvt} and \ref{n1gvt} reveal the existence of a number of unstable modes that 
are present in the disk and that dominate for different ranges of thermal relaxation times, for representative
combinations of the radial and vertical wavenumbers. In what follows, we shall identify the different unstable 

In order to accomplish this task, we work with reduced dispersion relations that we obtain by 
approximating the general dispersion relation, Equation (\ref{fulldisprel}), with the aim of describing the unstable modes that we identify. 
The ultimate justification for our approach is the good agreement 
that we find between the solutions that result from the full dispersion relation with those obtained from their reduced versions.
While necessarily crude, this procedure offers insights into the key physical ingredients driving each of the
unstable modes that we identify, as we discuss in detail below.

\subsection{(Magneto-) Acoustic Overstability}
\label{subsec_aco}

\begin{figure*}
\centering
\includegraphics[width=0.495\textwidth]{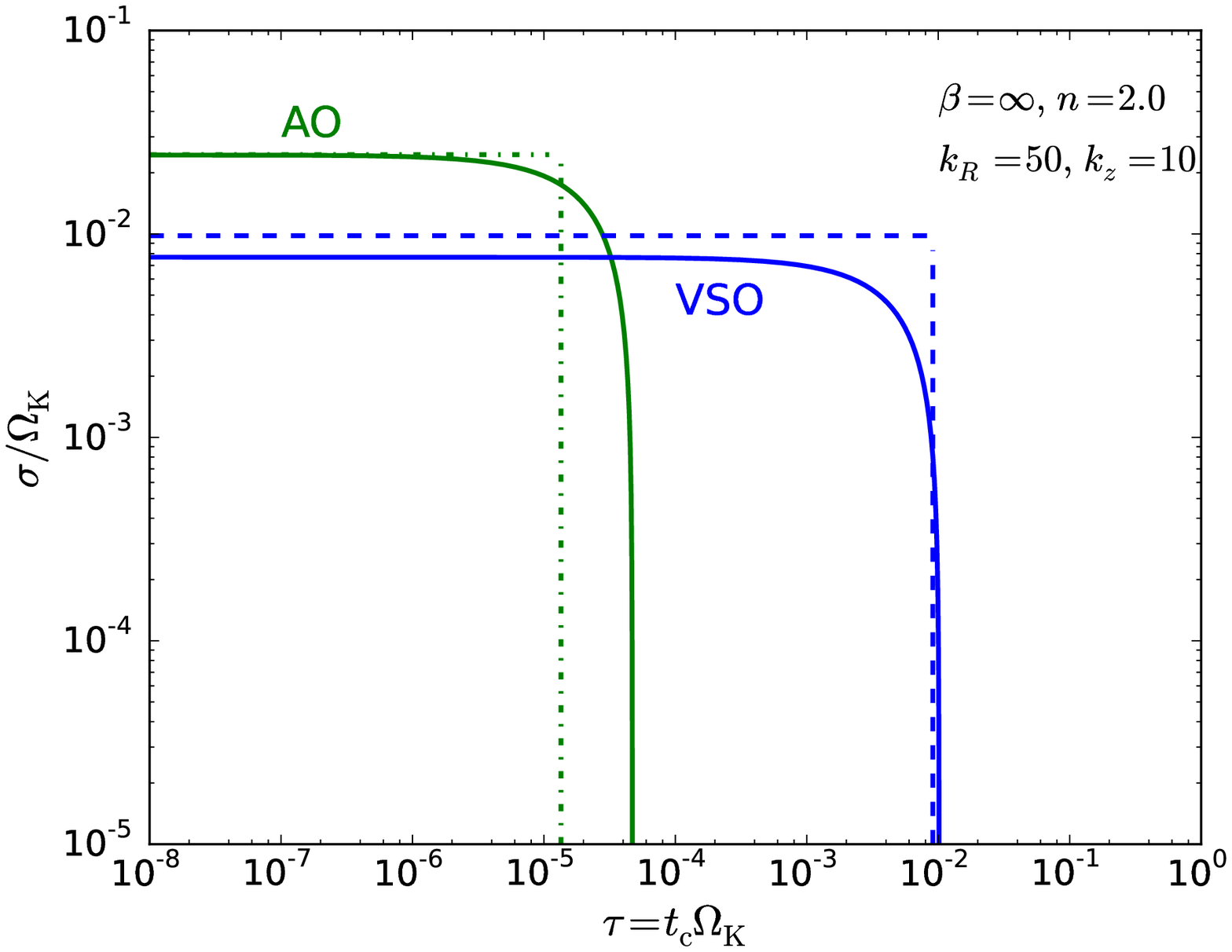}
\includegraphics[width=0.495\textwidth]{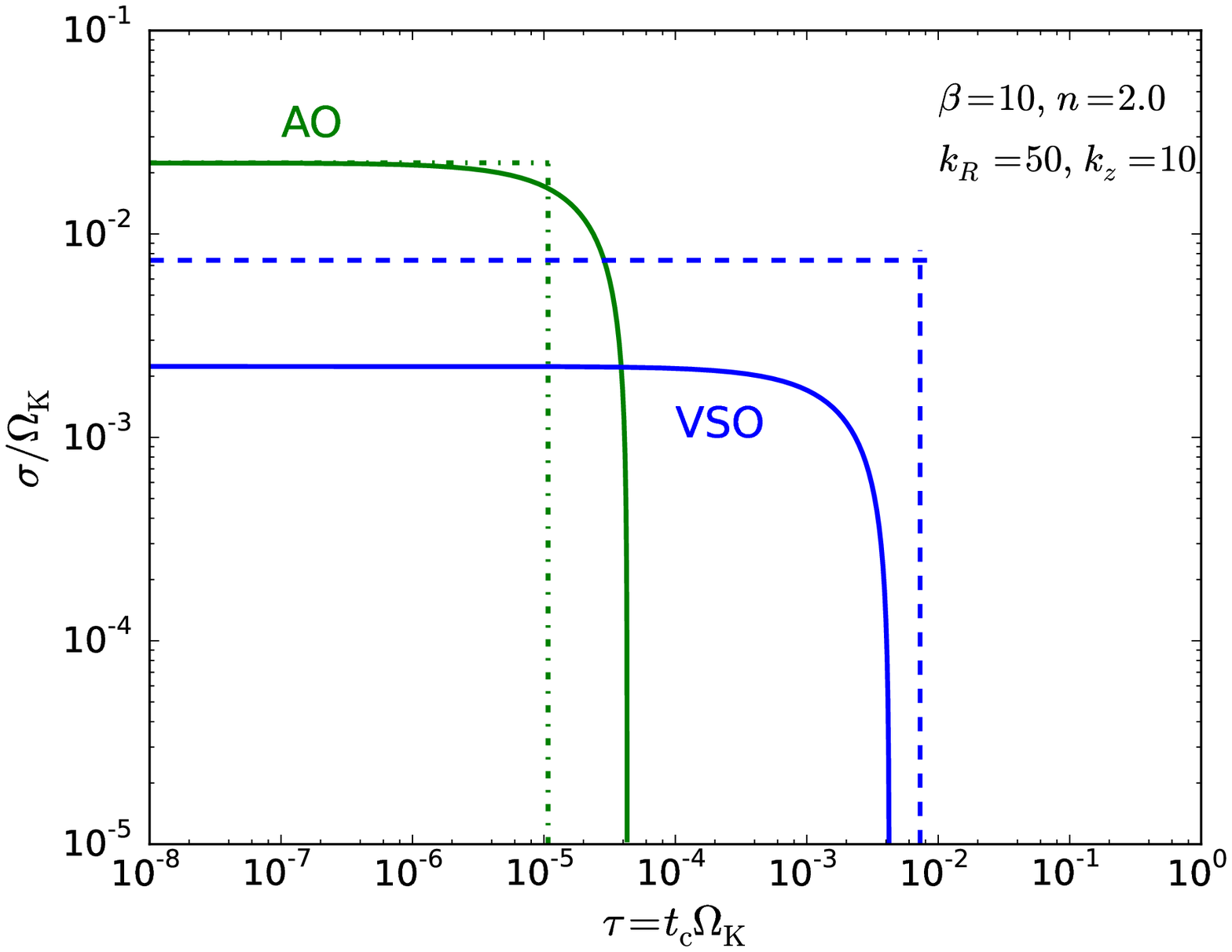}   \\
\caption{Growth rate of the unstable mode versus thermal relaxation time. Inclusion of vertical gradients and in particular vertical shear gives rise to a different overstable mode. The disk model parameters used are $(p, q, h, \gamma) = (-2, -1, 0.05, 1.4)$. The unstable growth rate $\sigma$ and the cooling time $\tau$ are non-dimensionalized in terms of $\Omega_{\rm K}$. The abbreviations used are: AO - Acoustic Overstability, VSO - vertical shear overstability \\ \\}
\label{n2gvt}
\end{figure*}

\begin{figure*}
\centering
\includegraphics[width=0.495\textwidth]{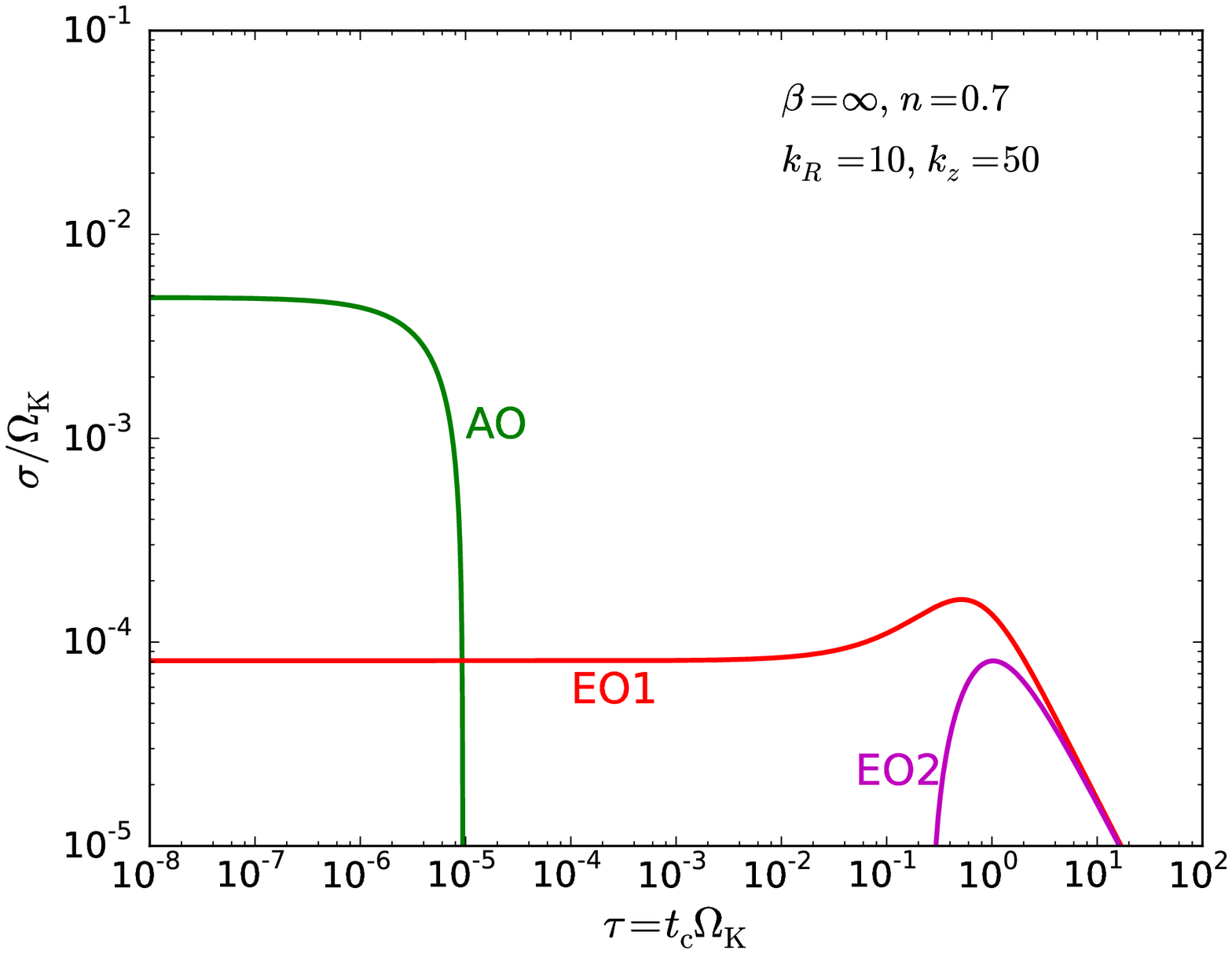}
\includegraphics[width=0.495\textwidth]{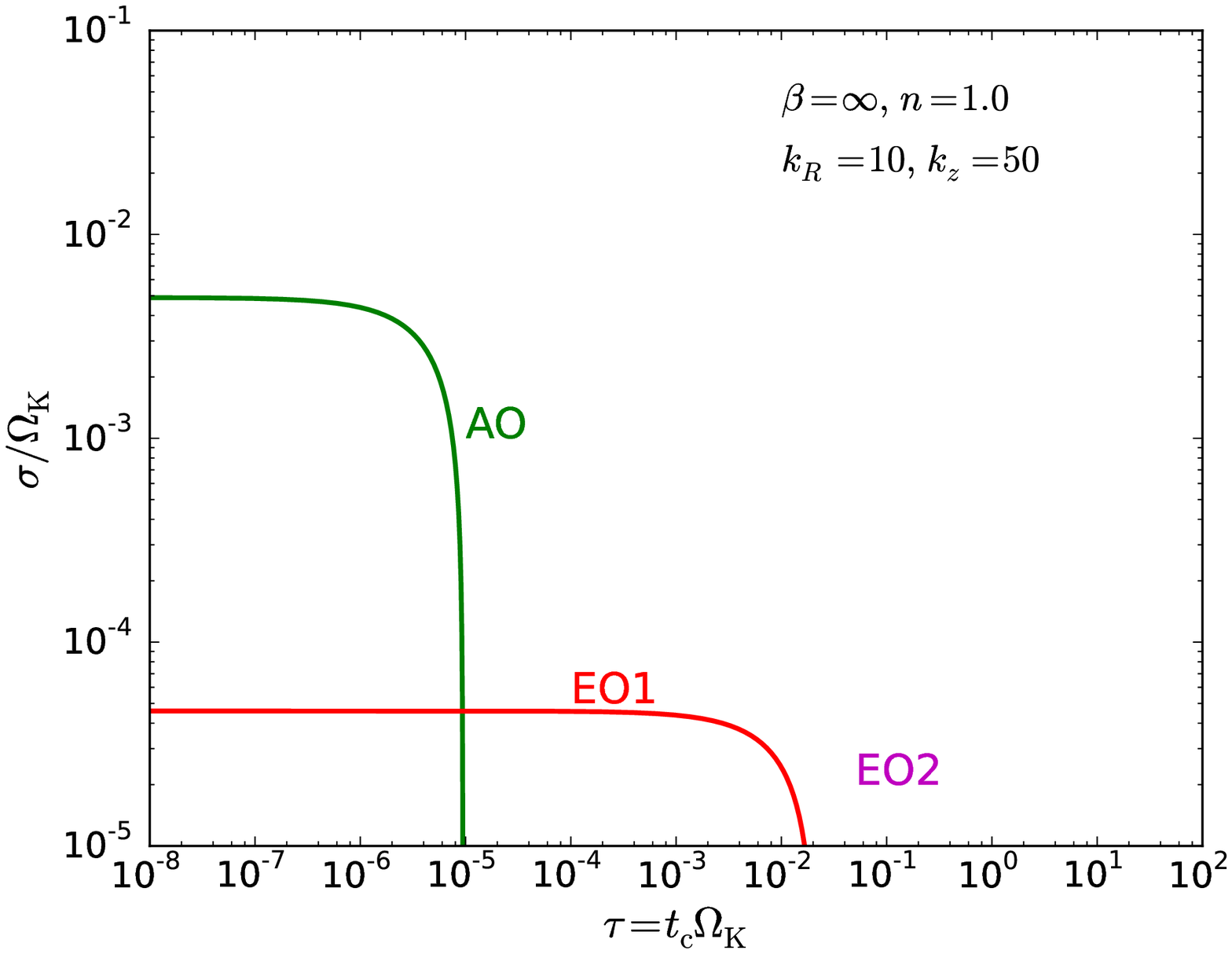}   \\
\caption{ Growth rate of the unstable mode versus thermal relaxation time. Illustration of the suppression of the epicyclic modes with increasing vertical shear. The disk model parameters used are $(p, q, h, \gamma) = (-2, -1, 0.05, 1.4)$. The unstable growth rate $\sigma$ and the cooling time $\tau$ are non-dimensionalized in terms of $\Omega_{\rm K}$. The abbreviations used are: AO - Acoustic Overstability, EO1 - first radial epicyclic Overstability and EO2 - second radial epicyclic overstability \\}
\label{n1gvt}
\end{figure*}

In Figures \ref{n0gvt}, \ref{n2gvt} and \ref{n1gvt}, we find an unstable mode, depicted by the solid green curve, that maintains an appreciable growth rate for very short cooling times and is swiftly damped beyond a critical time scale. 
This mode has an associated oscillatory part with a frequency, $\nu = Re[{\omega}]$, corresponding to that of acoustic waves in the disk. We, therefore, associate this mode with a rapidly oscillating overstable (magneto-)acoustic mode. 
The eigenvector of this acoustic overstable mode is shown in Figure \ref{eig:1}.

The acoustic overstability is only present for very short cooling times. 
In order to capture the basic characteristic features of this mode, we 
reduce the dispersion relation (\ref{fulldisprel}) to the simpler form

\begin{align}
\label{aodisp}
\omega^2 + i t_{\rm c}  F^{\rm A}_3 \omega -  F^{\rm A}_4 = 0 \,,
\end{align}
with
\begin{align}
 F^{\rm A}_3 & =  \gamma k^2 c_{\rm i}^2 \left( 1 + \frac{2}{\gamma \beta} \right) \,, \\
 F^{\rm A}_4 & = k^2 c_{\rm i}^2   \left( 1 + \frac{2}{ \beta} \right) + i k_R \frac{\partial c_{\rm i}^2}{\partial R} \, .
\end{align}
The coefficients $F^{\rm A}_3$ and $F^{\rm A}_4$ are reduced forms of the coeffcients $F_3$ and $F_4$ given by Equations (\ref{c_f3}) and (\ref{c_f4}). Here we have neglected terms containing shear and background pressure or density gradients from the coefficients in order to keep the analysis manageable yet retain the terms essential for instability. Notice that Equation (\ref{aodisp}) reduces to the quadratic form of the locally isothermal dispersion relation obtained when considering the limit, $t_{\rm c} \rightarrow 0$, in Equation (\ref{fulldisprel}). Substituting $\omega = \nu + i \sigma$, we can express the unstable growth rate as
\begin{align}
\label{aogr}
\sigma \sim k_R \frac{\partial c_{\rm i}^2}{\partial R} \left[ 4 k^2 c_{\rm i}^2  \left( 1 + \frac{2}{\gamma \beta} \right) \right]^{-1/2} \,.
\end{align}
The critical cooling time beyond which the overstable mode is rapidly damped is approximately
\begin{align}
\label{aotc}
t_{\rm c, cr} \sim \sigma / F^{\rm A}_3 \,.
\end{align}
The dash-dotted green colored lines in Figures \ref{n0gvt} and \ref{n2gvt} represent the numerical estimates for 
the growth rates and time scales obtained from Equations (\ref{aogr}) and (\ref{aotc}). Given the simplifications involved in obtaining these estimates, 
the agreement between the predictions for the growth rate and critical timescale from the analytical approximation and the numerical result is quite remarkable. 
In a magnetized disk, it is the fast magnetoacoustic waves that undergo overstable oscillations. The field itself has a very small effect on the growth rate and the tendency is to damp the amplification of the oscillations. 

The thermal gradient is solely responsible for amplifying the otherwise stable acoustic waves. The thermal gradient adds to the restoring forces in such a way as to cause the amplitude of the near isothermal disturbances to grow exponentially with each oscillation.

\subsection{Unstable Epicyclic Oscillations and the Convective Overstability}
\label{subsec_ep}

When the vertical structure of the disk is ignored, we find two overstable modes that are quite similar in character. 
Figure \ref{n0gvt} shows two modes with peak growth rates around cooling time, 
$\tau \approx 1$. They have the exact same growth rates for $\tau > 1$ but acquire different values for shorter cooling times, i.e., $\tau < 1$.  Interestingly, these modes are reminiscent of a recently discovered unstable mode that has been termed the convective overstability. 

\begin{figure*}
\centering
\subfigure[First Epicyclic Mode]{
\includegraphics[width=0.45\textwidth]{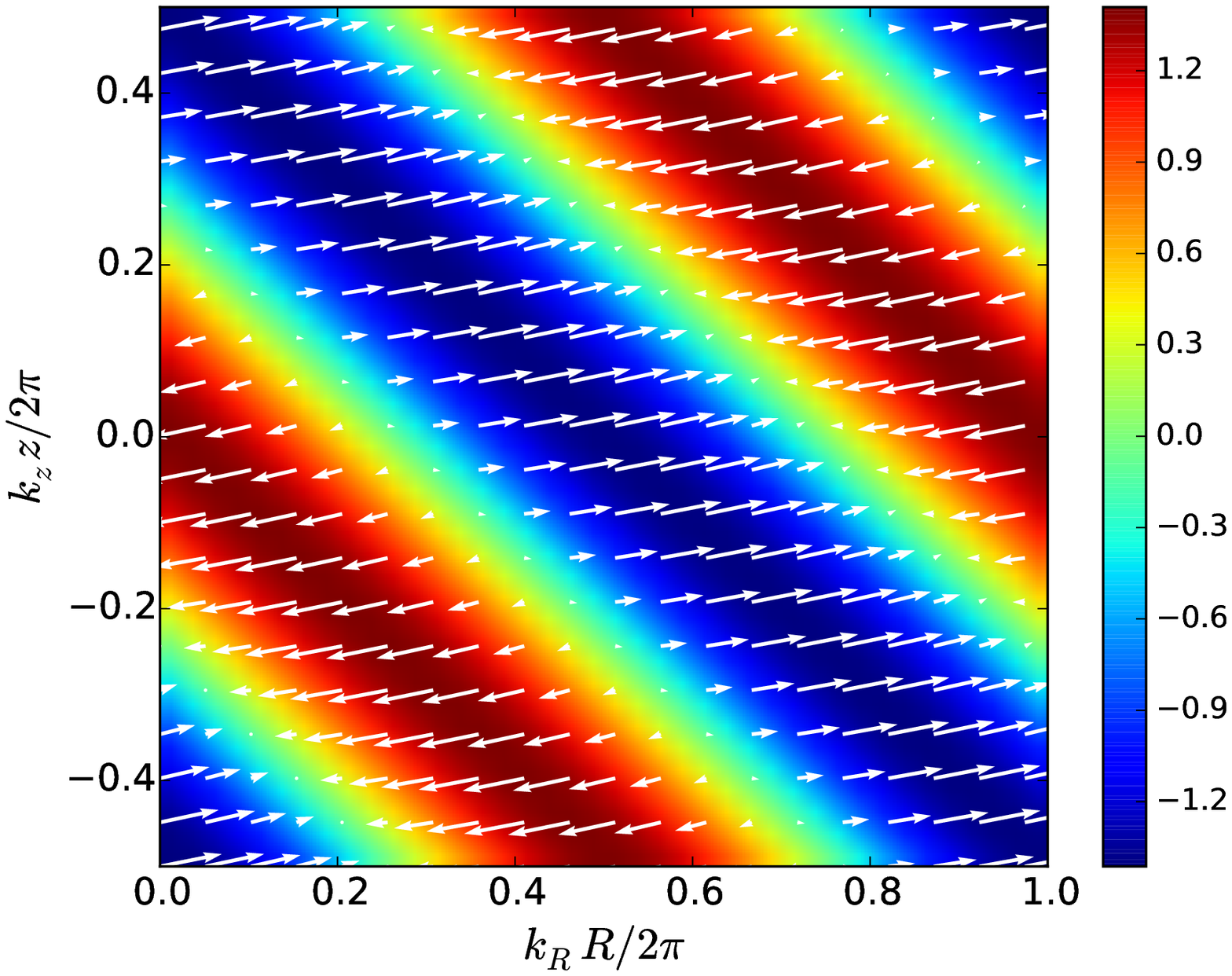}\label{eig:1}}
\subfigure[Second Epicyclic Mode]{
\includegraphics[width=0.45\textwidth]{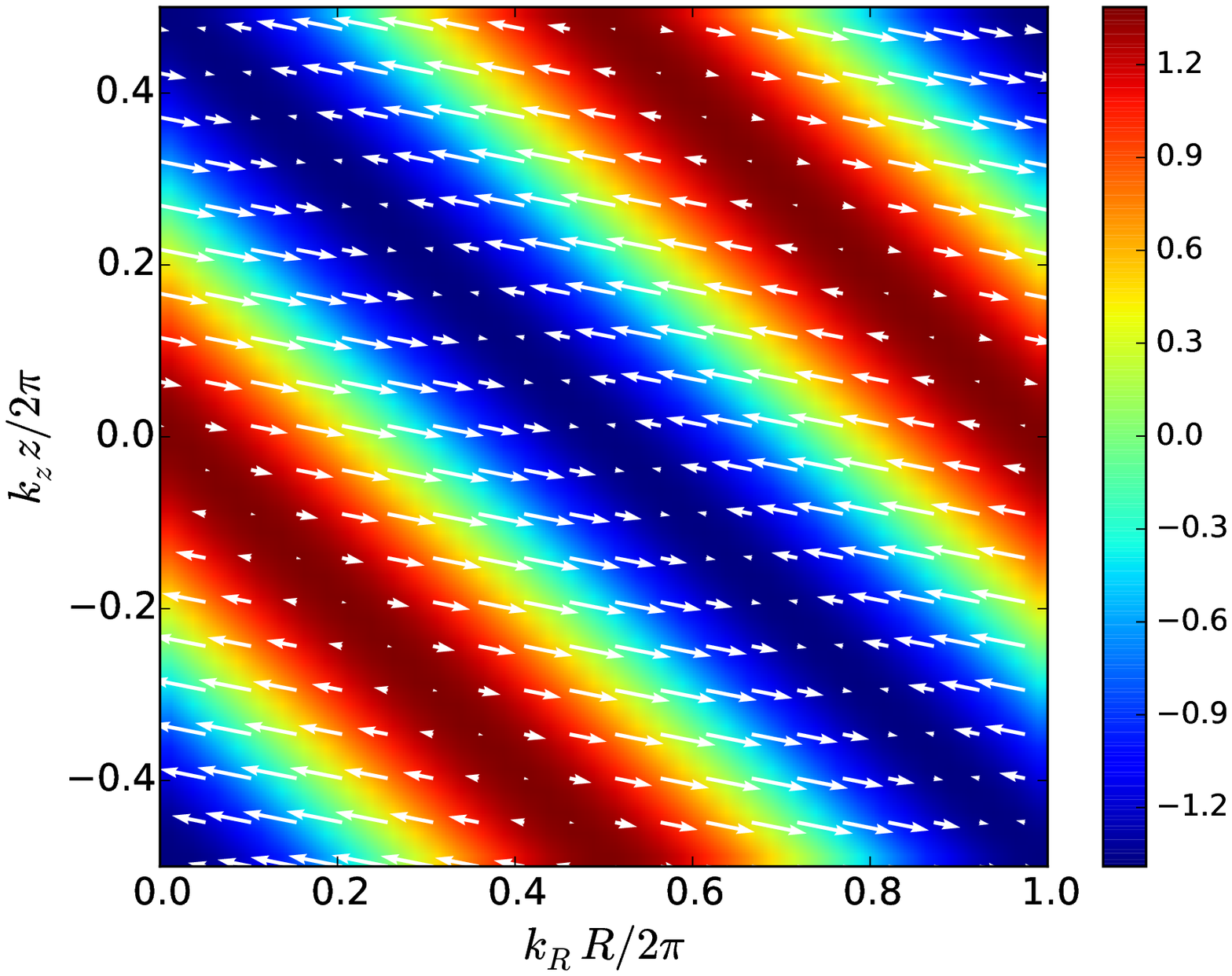}\label{eig:2}}
\subfigure[ Acoustic Mode]{
\includegraphics[width=0.45\textwidth]{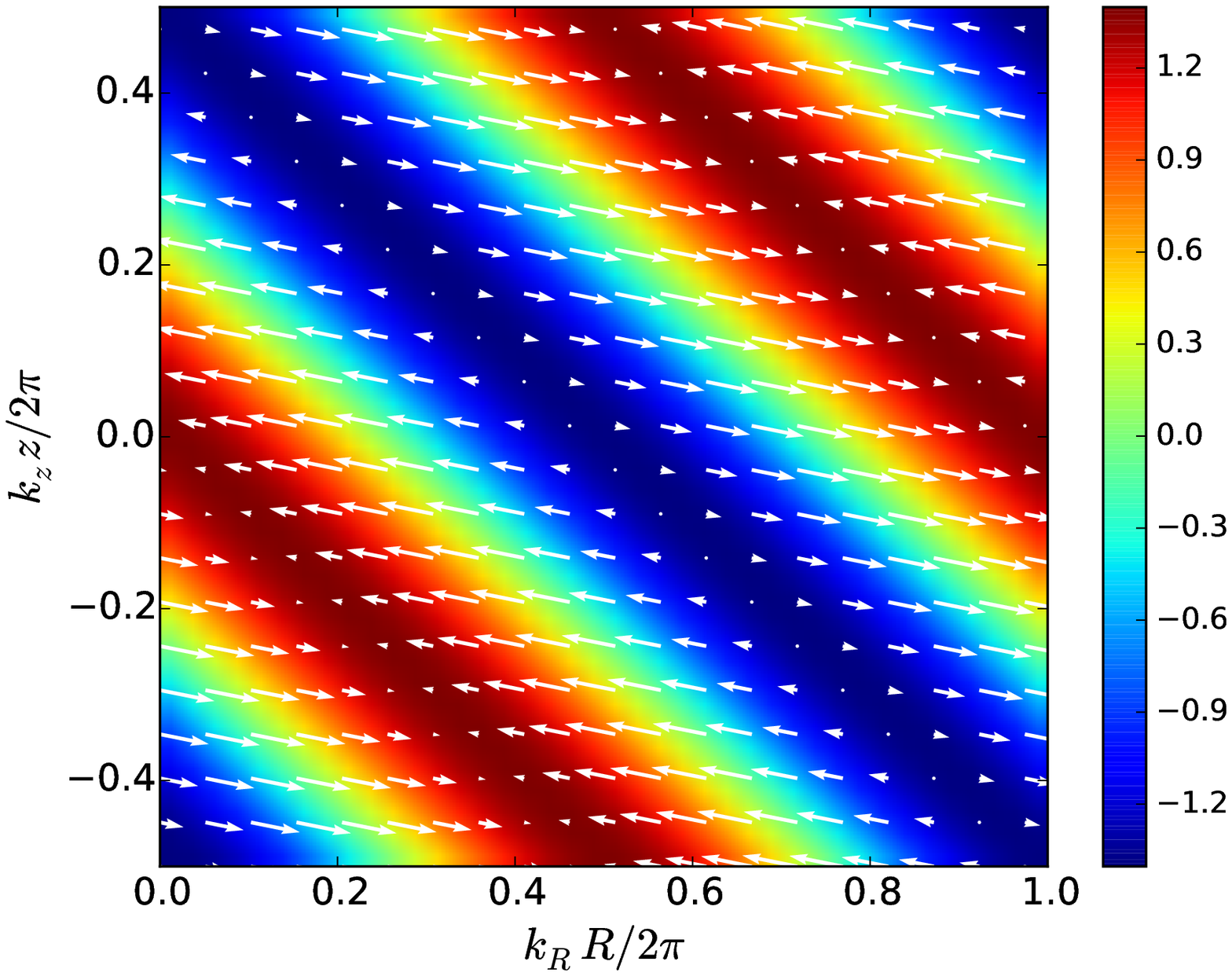}\label{eig:3}}
\subfigure[Vertical Shear Mode]{
\includegraphics[width=0.45\textwidth]{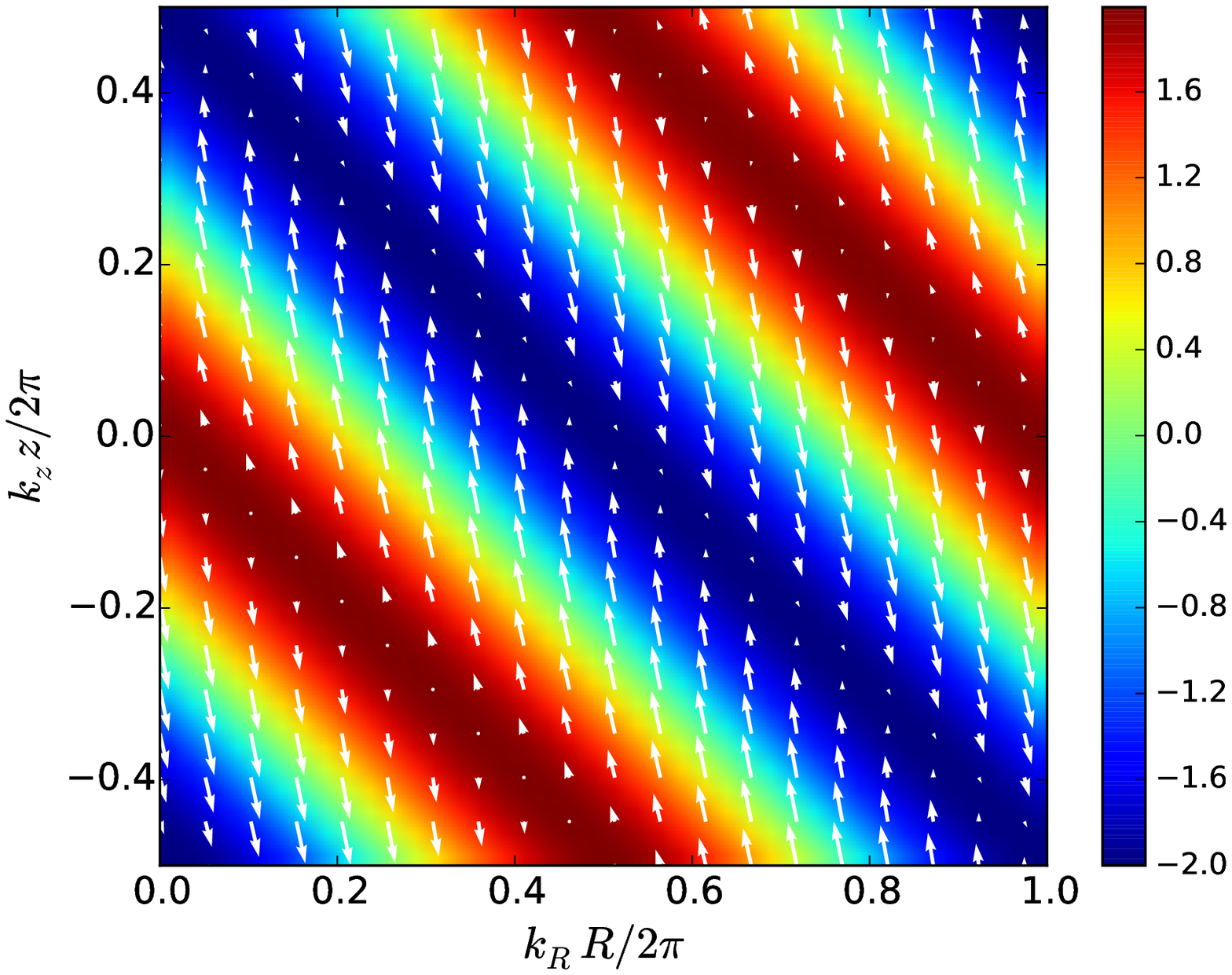}\label{eig:4}}  
\caption{ Illustration of the unstable eigenvectors. The colored isocontours represent density perturbations and the arrows represent the flow in the $R-z$ plane. (a) and (b) represent the first and second epicyclic overstable mode respectively with $k_R H = 10$ and $k_z H = 50$ , (c) is the acoustic overstability with $k_R H = 50$ and $k_z H = 10$ and (d) the vertical shear overstable mode with $k_R H = 50$ and $k_z H = 10$. The disk model parameters used are $(p, q, h, \gamma, \beta) = (-2, -1, 0.05, 1.4, \infty)$.}
\label{eigmodes}
\end{figure*}

When considering disks without vertical structure, \citet{kh14, lyra14} find that an unstable mode is present if the radial Brunt-V\"{a}is\"{a}la frequency of the system is negative, which corresponds to the familiar Schwarzschild criterion for convective instability to adiabatic perturbations. However, provided $k_z \gg k_R$, the overstable mode has significant growth rates for a small range of thermal relaxation times centred around, $\tau \approx 1$. For comparison, we plot in Figure \ref{n0gvt} the growth rate of the convective overstable mode by solving the dispersion relation in (\citealt{lyra14}, equation 25) for the same set of model parameters used here. It is apparent that there is a striking correspondence between the two nearly degenerate modes, that we see represented by the solid red and magenta curves in Figure \ref{n0gvt}, with the convective overstability represented by the dashed cyan curve.
For ease of reference and to avoid confusion, we shall henceforth refer to the solid red curve as the \emph{first} epicyclic mode and the solid magenta curve as the \emph{second} epicyclic mode.Examination of the unstable eigenvalue and eigenvector, see Figures \ref{eig:1} and \ref{eig:2}, reveals 
that these modes comprise of growing epicyclic oscillations. This is also the basic nature of the convective overstable modes as described by \citet{kh14, lyra14}.

Figure \ref{n0gvt} also shows that the first epicyclic mode is present for shorter thermal relaxation times analogous to the acoustic overstable mode discussed in the previous section. In the locally isothermal limit, $t_{\rm c} \rightarrow 0$, the dispersion relation, Equation (\ref{fulldisprel}) becomes analytically tractable as it is in the locally adiabatic limit. When vertical structure is ignored, we find that two distinct overstable modes are present in this limit of infinitely fast thermal relaxation. One of them is the aforementioned overstable (magneto-)acoustic oscillations. The other mode is one in which the radial temperature gradient acts to drive epicyclic oscillations unstable. It is this other unstable mode that we observe as the first epicyclic mode in Figure \ref{n0gvt}. 

We derive below characteristic growth rates and time scales associated with the first epicyclic mode. 
Performing a similar approximate analysis as in the previous subsection, we can derive analytical expressions that characterize the mode that persists at shorter cooling times.  In the short cooling time limit, we obtain the reduced dispersion relation
\begin{align}
\label{eo1disp}
\omega^2 + i  t_{\rm c} F^{\rm E}_5 \omega -  F^{\rm E}_6 = 0,
\end{align}
with
\begin{align}
 F^{\rm E}_5 & = \gamma \mu^2  (\kappa_R^2 + N_R^2)   \left[ 1 - \frac{i k_R}{k^2 \left( 1 + 2/\beta \right)}\frac{\partial \ln c_{\rm i}^2}{\partial R}  \right]      ,\\
 F^{\rm E}_6 & = \mu^2 \kappa_R^2 \left[ 1 - \frac{i k_R}{k^2  \left( 1 + 2/\beta \right)}\frac{\partial \ln c_{\rm i}^2}{\partial R}  \right] \,,
\end{align}
where we have neglected the vertical background gradients. The growth rate is then approximately
\begin{align}
\label{eogr}
\sigma \simeq - \frac{ \kappa_R k_R k_z}{\sqrt{2} k^3 \left( 1 + 2/\beta \right)}\frac{\partial \ln c_{\rm i}^2}{\partial R}   \, .
\end{align}
Were it not for the imaginary term containing the radial temperature gradient, this solution would simply represent stable epicyclic oscillations. The estimate for the growth rate (represented by the dotted red line) given by Equation (\ref{eogr}) agrees remarkably well with the numerical result as seen in Figure \ref{n0gvt}. 
Continuing with the approximate analysis, we extract a characteristic timescale at which the growth rate declines rapidly
\begin{align}
\label{eotc}
t_{\rm c, cr} \sim \sigma /  F^{\rm E}_5 .
\end{align}

Although derived assuming very small cooling times, the critical time 
expressed by Equation (\ref{eotc})  coincides with the timescale for maximum 
growth provided that, $k_R \ll k_z$.  While the growth rate does in fact fall  
beyond this critical time scale, it does not do so abruptly when an unstable
radial entropy gradient is present. 

The first epicyclic mode constitutes overstable epicyclic oscillations that are also
referred to as \emph{acoustic-inertial} oscillations. Isothermal perturbations
in the presence of a radial temperature gradient add to the restoring forces as a 
fluid element is perturbed radially leading to overstability. 

The second epicyclic overstable mode plotted in Figure \ref{n0gvt} bears a much closer resemblance to the dashed cyan curve representing the convective overstability of \citet{kh14, lyra14}. 
\citet{lyra14} performed a compressible local linear mode analysis of the convective overstability and derived analytical expressions for the characteristic growth rate and time scales.  The dispersion relation given by Equation (\ref{fulldisprel}) in Section \ref{sec_dispersionrelation}, reduces to Equation (17) in \citet{lyra14} if we set all vertical 
gradients and the magnetic field to zero. However, the radial temperature gradient was omitted from the dispersion relation, Equation (19) in \citet{lyra14} where it was argued that this term may be ignored as it is proportional to $1/R$. This exclusion cannot be rigorously justified and results in the divergent behaviour of the epicyclic modes remaining unseen. 
One possible argument to ignore the temperature gradient is to restrict the analysis to modes with $k_R = 0$. However, this would go against the spirit of the short wave approximation upon which the entire local analysis is predicated. At the same time, it would also lead to growing modes that are essentially scale-independent. Ultimately, the justification for omitting any term ought to be consolidated by comparing against numerical solutions. Our results indicate that retaining the term containing the temperature gradient has a fundamental effect on the nature of the modes as illustrated by Figure \ref{n0gvt} and should therefore not be discarded.

Despite these issues, the calculations by \citet{lyra14} present the closest one can get to describing the second epicyclic mode analytically. Including the radial temperature gradient renders the dispersion relation intractable to analysis and it becomes difficult to extract even approximate expressions that characterize the unstable growth rate and time scales. Therefore, we rely on the analysis in \citet{lyra14} and generalize it to include a toroidal magnetic field. 

If we work in the anelastic approximation ($c_{\rm i} \rightarrow \infty$), as in \citet{lyra14}, the dispersion relation (\ref{fulldisprel}) assumes the form 
\begin{align}
\label{codisprel}
\omega^3 + i \varsigma \varepsilon \omega^2 - \omega \left[ \kappa_R^2 + \bar{M}_R^2 \right] - i \varsigma \mu^2 \left[ \kappa_R^2 \varepsilon + \bar{L}_R^2  \right] = 0,
\end{align}
where we have made use of the definitions\footnote{We adopt the definitions used in \citet{lyra14}
to facilitate the comparison.}
\begin{align}
\mu^2 &\equiv \frac{k_z^2}{k^2},   
\end{align}
\begin{align}
\varsigma &\equiv \frac{1}{\gamma t_{\rm c}} ,   \\
\varepsilon &\equiv \frac{(1 + 2/\beta)}{(1 + 2/\gamma \beta)} ,  \\
L_R^2 &\equiv - \frac{c_{\rm A}^2}{c_{\rm s}^2}\frac{1}{\rho_0}\frac{\partial P_T}{\partial R}\frac{\partial}{\partial R}\ln \left( \frac{B}{\rho} \right),   \\
\bar{L}_R^2 &\equiv \frac{L_R^2}{(1 + 2/\gamma \beta)} , \\
\bar{M}_R^2 &\equiv \frac{(N_R^2 + L_R^2)}{(1 + 2/\gamma \beta)}  \,. 
\end{align}

In order to find approximate solutions to the dispersion relation (\ref{codisprel}), it is convenient to express $\omega$ explicitly as a complex variable, i.e., $\omega = \nu + i \sigma$,
in the dispersion relation, Equation (\ref{codisprel}). The dispersion relation now splits it into a real and imaginary part both of which must equal to zero, respectively,
\begin{eqnarray}
\label{mconoreal}
\nu^2 &-& \mu^2 \left[ \kappa^2 + \bar{M}_R^2 \right] - 3 \sigma^2 - 2 \sigma \varepsilon =0 ,  \\
\label{mconoimag}
\sigma^3 &+& \varsigma \varepsilon \sigma^2 - \sigma \left[ 3 \nu^2 - \mu^2 \bar{M}_R^2 \right]    \nonumber \\
&-& \varsigma \left\{ \nu^2 - \mu^2 \left[ \varepsilon \kappa_R^2 + \bar{L}_R^2  \right]  \right\} = 0.
\end{eqnarray}
Substituting Equation (\ref{mconoreal}) in Equation (\ref{mconoimag}) yields
\begin{align}
8 \sigma^3 + 8 \sigma^2 \varsigma + 2 \sigma \left[ \varsigma^2 + \mu^2 \left( \kappa^2 + \bar{M}_R^2 \right) \right] \nonumber \\ 
+ \varsigma \mu^2 \left[ \varepsilon \bar{N}_R^2 + (\varepsilon - 1) \bar{L}_R^2 \right]  = 0.
\end{align}
We can obtain an approximate solution for this equation by noting that
$\epsilon \simeq 1$ except for very small values of $\beta$ ($\varepsilon = 1.23$ with 
$\beta = 1$ and $\gamma = 1.4$). Since $L_R^2 \ll N_R^2$, except for very strong fields, 
in the limit of $\sigma \ll \varsigma$, we can neglect the cubic and quadratic terms, to obtain
\begin{align}
\label{cogr}
\sigma \simeq -\frac{\varsigma \mu^2 \bar{N}_R^2}{2 \left\{ \varsigma^2 + \mu^2 \left[ \kappa^2  + \bar{M}_R^2  \right] \right\}}.
\end{align}
The criterion for this convective overstability is still effectively $N_R^2 < 0$ and the maximum growth rate 
\begin{align}
\label{cogrmax}
\sigma_{\rm max} = - \frac{ |\mu| \bar{N}_R^2 }{4 \sqrt{  \kappa^2  + \bar{M}_R^2  }} \,,
\end{align}
is achieved for
\begin{align}
\label{cotm}
t_{\rm {c, max}} = \frac{1}{\sqrt{ \gamma^2 \mu^2 \left[ \kappa^2 + \bar{M}_R^2 \right]}} \,.
\end{align}

The effect of the magnetic field on the first epicyclic mode is to marginally dampen the growth rate. The 
growth rate of the second epicyclic mode appears slightly enhanced around
$\tau \simeq 1$ for $\beta = 10$. This is because the magnetic contribution to pressure support becomes
larger for stronger fields and hence the effective radial gravitational acceleration increases. This causes the
Brunt V\"{a}is\"{a}la frequency, in the manner we have defined it, to increase for stronger fields and consequently
leads to a slight increase in growth rate.

\subsection{Vertical Shear Overstability}
\label{subsec_vso}

When the vertical gradients of the background disk are taken into account, we identify another fundamental mode of instability, which is illustrated as the solid blue curve in Figure \ref{n2gvt}. As we shall see below, this mode can be attributed to 
the destabilizing effect of vertical shear in the disk. 

Instability due to vertical shear has received much attention of late in connection with protoplanetary disks \citep{ngu13, cmmp14, bl15, ly15}. The destabilizing effect of vertical shear was first studied by \citet{gs67, fr68} and its role as a potential source of angular momentum transport in accretion disks was suggested by \citet{vuab98, vu03}. 

This unstable mode constitutes one of three overstable roots (along with the acoustic and acoustic-inertial modes) of the dispersion relation, Equation (\ref{fulldisprel}), in the locally isothermal limit, $t_{\rm c} \rightarrow 0$. 
Using indications obtained from the numerical solutions, we attempt to derive analytical approximations for the growth rate and critical timescale of the unstable mode due to vertical shear.
By retaining only the terms essential to instability as done in the previous sections, we obtain a reduced dispersion relation in the short cooling time limit 
\begin{align}
\label{vsodisp}
\omega^2 + i t_{\rm c}   F^{\rm V}_5 \omega -  F^{\rm V}_6 = 0,
\end{align}
with
\begin{align}
 F^{\rm V}_5 & = \frac{\gamma k_{R}^2 M_z^2}{ k^2 (1+2/\beta)},  \\
 F^{\rm V}_6 & = \kappa_R^2 \mu^2 - \frac{i \kappa_z^2 k_R}{k^2 c_{\rm i}^2 (1+2/\beta)} \frac{1}{\rho}\frac{\partial P_{\rm T}}{\partial z}.
\end{align}
Modes that are vertically elongated and radially narrow are known to tap the free energy in vertical shear more efficiently. For this reason, we have assumed $k_R \gg k_z$ in deriving the reduced form of the dispersion relation above. The unstable growth rate may then be approximated as
\begin{align}
\label{vsogr}
\sigma \sim \frac{ \kappa_z^2 k_R }{k^2 c_{\rm i}^2 (1+2/\beta) }  \frac{1}{ \sqrt{2} \kappa_R \mu \rho} \frac{\partial P_{\rm T}}{\partial z}  \,,
\end{align}
and the critical cooling time beyond which the mode is rapidly damped is
\begin{align}
\label{vsot1}
t_{\rm c, cr} \sim \sigma / F^{\rm V}_5 \,.
\end{align}
In the hydrodynamic limit, $\beta \rightarrow \infty$, and assuming vertical isothermality, the cooling time varies as
\begin{align}
\label{vsotc}
t_{\rm c, cr} &\sim \frac{\sigma}{N_z^2}   \propto \frac{1}{\gamma - 1} \,,
\end{align}
Equation (\ref{vsotc}) is qualitatively similar to the critical cooling time estimate derived by \citet{ly15} in the radially local vertically global setting. 
The approximate estimates of the growth rate and critical cooling time have been over-plotted onto Figure \ref{n2gvt}.
We find good agreement between the simplified estimates and the numerical values which implies that the basic ingredients of the instability are 
captured by the reduced analytical approximations.  
These modes are essentially overstable vertical epicyclic oscillations as is also suggested by the unstable eigenvector illustrated in Figure \ref{eig:4}. 

The unstable mode due to vertical shear that we have identified here has a different character to that which has been derived by linear analysis in previous studies.
The unstable mode in \citet{vuab98, vu03} is a pure instability that is determined by the criterion
\begin{align}
\kappa_R^2 - \kappa_z^2 \frac{k_R}{k_z} < 0,
\end{align}
which is satisfied only if $k_R$ and $k_z$ are of opposite sign.  The unstable mode due to vertical shear that we consider here under more general circumstances, is an overstable mode. In the isothermal limit $t_c \rightarrow 0$, the dispersion relation (\ref{fulldisprel}) reduces to Equation (32) derived in \citet{ngu13} if the radial background gradients are ignored and if we make the transformation $g \rightarrow g/2$, where $g$ is the vertical component of the gravitational acceleration. Note, however, that the crucial imaginary term in  $F^{\rm V}_6$ is missing in Equation (32) of \citet{ngu13}. We posit that this term is the real destabilizing source driving unstable motions in the disk. The fact that these unstable modes manifest as overstable oscillations is also borne out by the semi-global eigenmode calculations conducted in \citet{ngu13, cmmp14, ly15}. \\

\subsection{Coexistence of Epicyclic and Vertical Shear Modes}
\label{subsec_coexistence}

The numerical solutions in Section \ref{subsec_ep} were obtained without including the
vertical gradients in the dispersion relation whereas those presented in Section \ref{subsec_vso} 
takes account of both radial and vertical gradients. 
The exclusion of vertical structure in Section \ref{subsec_ep} was performed with the intention 
to relate our results to previous studies where the analysis was similarly restricted to the horizontal plane. 
The question arises as to whether the overstable epicyclic modes occur in the presence of vertical stratification. 
Figure \ref{n2gvt} does not contain the overstable radial epicyclic modes of Section \ref{subsec_ep} 
and seems to suggest that they are not present in a disk with vertical shear.
We attempt below to analytically describe the observed trend by expanding the approximate analysis of the
previous sections to include the terms relevant to the epicyclic and vertical shear modes
in the reduced dispersion relation
\begin{align}
\omega^2 + i \gamma t_{\rm c}  F^{EV}_5 \omega - F^{EV}_6 = 0,
\end{align}
with
\begin{align}
F^{EV}_5 & =  \mu^2 \kappa_R^2 \left\{ 1 - \frac{i k_R}{k^2 \left( 1 + 2/\beta \right)}\frac{\partial \ln c_{\rm i}^2}{\partial R}  \right\} ,  \\
F^{EV}_6 & = \left[  \mu^2 \kappa_R^2 - \frac{i \kappa_z^2 k_R}{k^2 c_{\rm i}^2 (1+2/\beta)} \frac{1}{\rho}\frac{\partial P_{\rm T}}{\partial z}   \right]  \nonumber \\
& \times  \left\{ 1 - \frac{i k_R}{k^2 \left( 1 + 2/\beta \right)}\frac{\partial \ln c_{\rm i}^2}{\partial R}  \right\} .
\end{align}
The growth rate in this general situation can be approximated as 
\begin{align}
\label{eovsogr}
\sigma \simeq \left| \frac{ \kappa_R k_R k_z}{\sqrt{2} k^3}\frac{\partial \ln c_{\rm i}^2}{\partial R} +  \frac{ \kappa_z^2 k_R }{k^2 c_{\rm i}^2 }  \frac{1}{ \sqrt{2} \kappa_R \mu \rho} \frac{\partial P_{\rm T}}{\partial z}    \right|
\end{align}
In the absence of vertical shear, the growth rate, given by Equation (\ref{eovsogr}), reduces to that of the radial epicyclic mode. When vertical shear is present, Equation (\ref{eovsogr}) suggests a competition of sorts between the radial epicyclic mode and vertical shear mode. For substantial vertical shear strengths and in particular for large radial wavenumbers, the overstable vertical shear mode alone survives. This is in accordance with the trend we observe from the numerical calculations. To illustrate the point, Figure \ref{n1gvt} shows gradually weakening radial epicyclic modes as the vertical shear strength is increased. 
Our results suggest that the overstable radial epicyclic modes may not play a significant role in realistic accretion disks. Further studies of these modes in a global context will
help to establish their relevance. 

\subsection{Scale Dependence of Unstable Modes}
\label{subsec_scale}

\begin{figure}
\centering
\label{sf:1}
\subfigure[First Epicyclic Mode]{
\includegraphics[width=4cm,height=3cm]{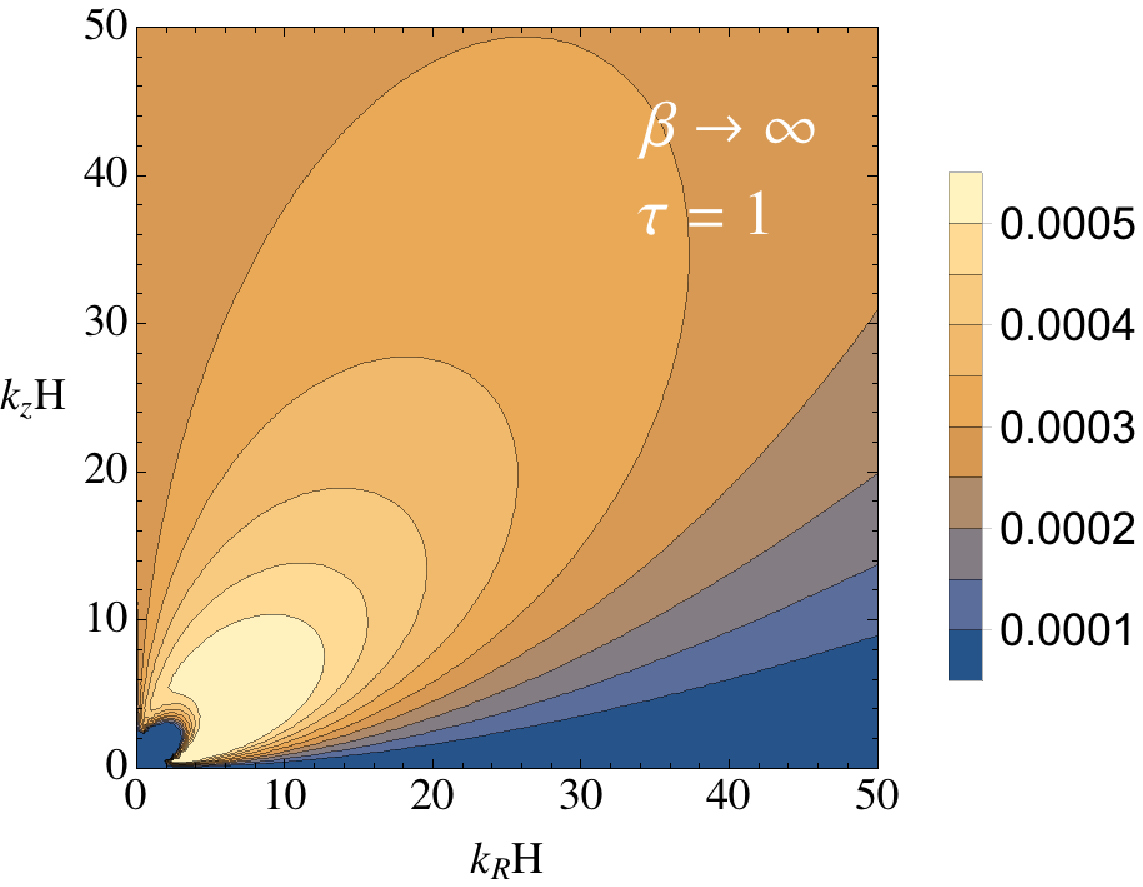}\label{cp:a}}
\subfigure[Second Epicyclic Mode]{
\includegraphics[width=4cm,height=3cm]{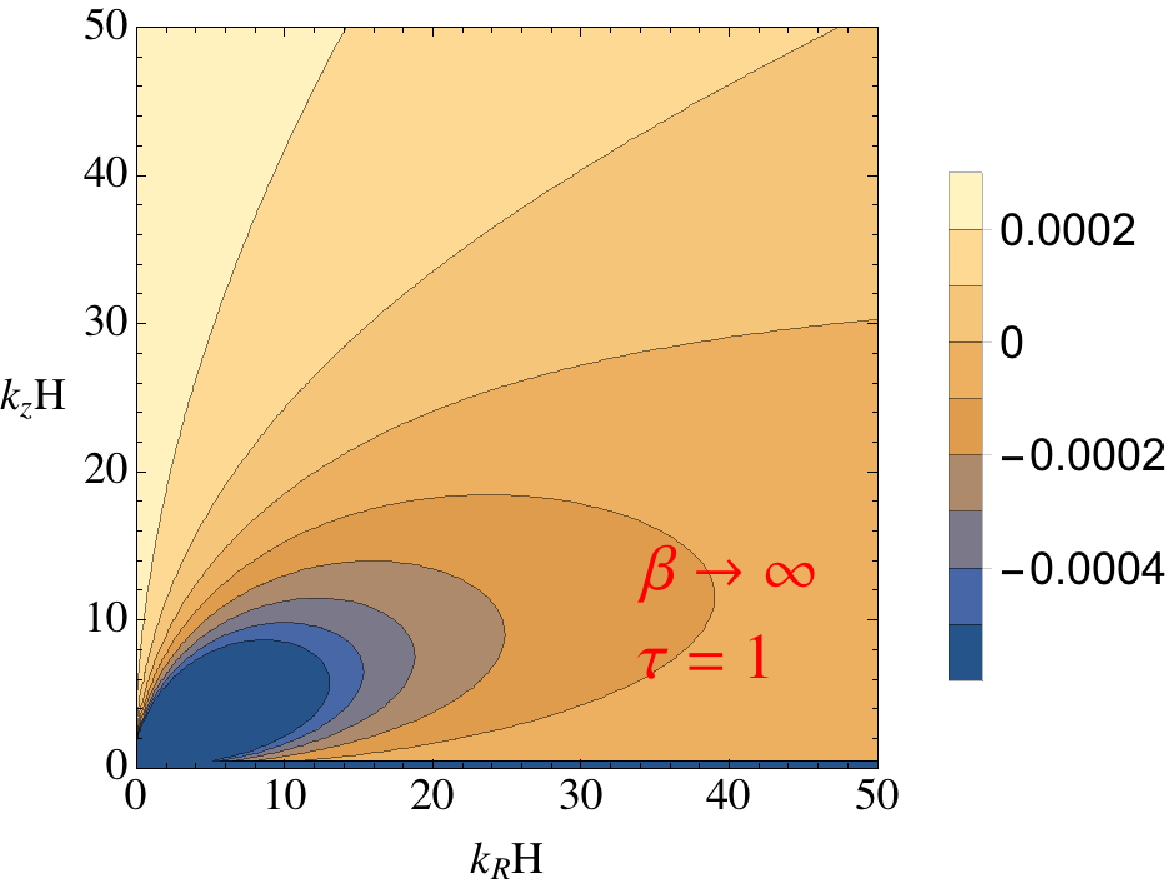}\label{cp:b}}  
\subfigure[Acoustic Mode]{
\includegraphics[width=4cm,height=3cm]{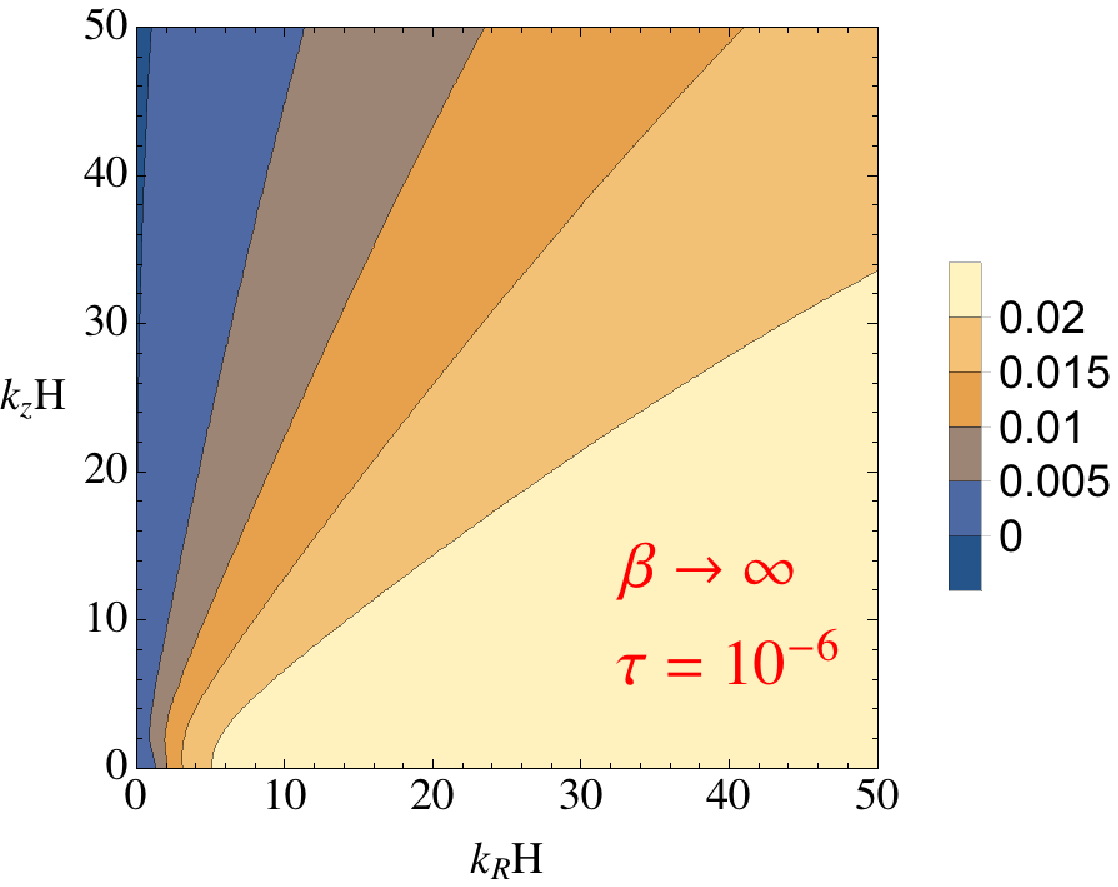}\label{cp:c}}
\subfigure[Vertical Shear Mode]{
\includegraphics[width=4cm,height=3cm]{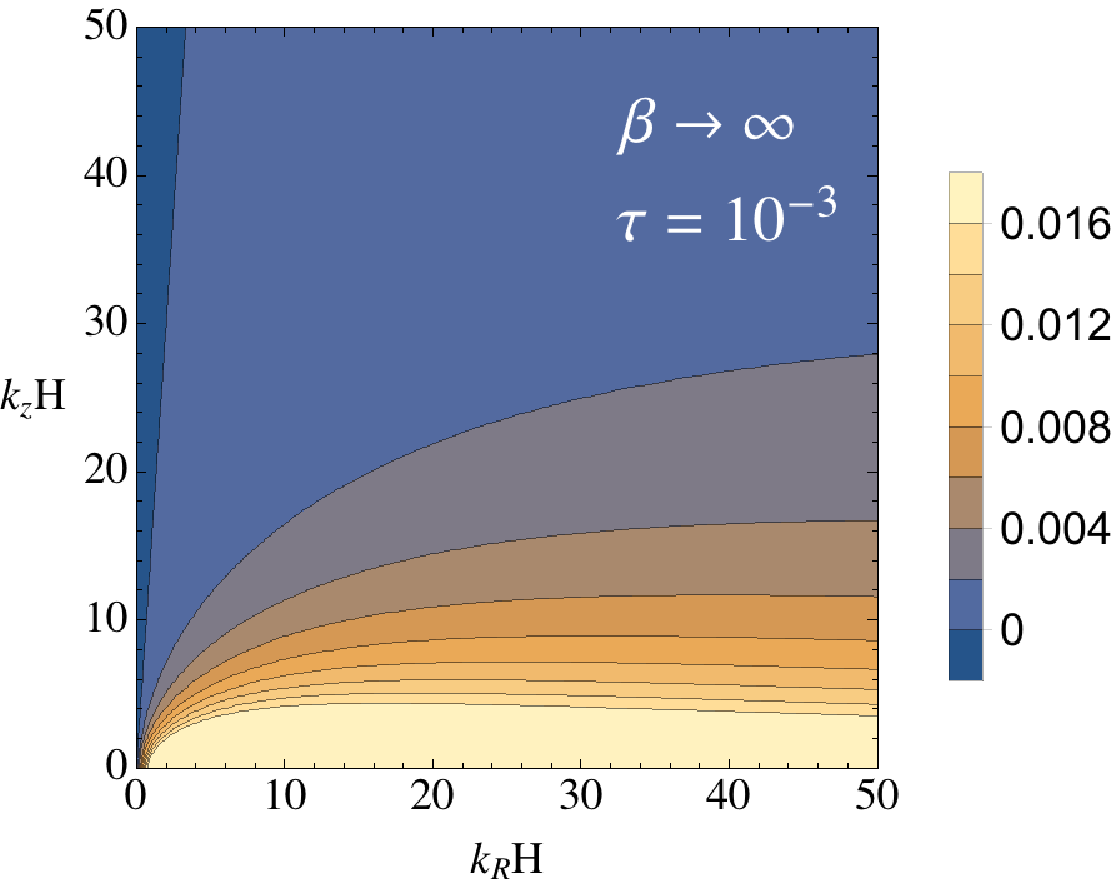}\label{cp:d}}
\caption{ The growth rates as a function of the perturbation wavenumbers for the disk model parameters $(p, q, h, \gamma) = (-1.5, -1, 0.05, 1.4)$. The coloured iso-contours represent the numerically computed values of the growth rates. (a) and (b) represent the first and second epicyclic overstable modes. (c) represents the acoustic overstable mode and (d) represents the vertical shear overstable mode. The preponderance of these modes stays the same for lower values of $\beta$.}
\label{fig:contplots}
\end{figure}

Our analysis thus far focussed on studying the unstable growthrates
and critical timescales as a function of the thermal relaxation time and magnetic field strength. 
This was done for fixed representative values of the perturbation wavenumbers. 
Here, we examine the length scale preferences
of the different modes identified in Figures \ref{n0gvt} and \ref{n2gvt}.
Figure \ref{fig:contplots} provides a graphical overview of the presence and strength of the 
unstable modes in wavenumber space obtained by numerically solving the general
 dispersion relation in Equation (\ref{fulldisprel}).

The acoustic overstable mode of Section \ref{subsec_aco} is seen to be dominant in the region of wavenumber space with $k_R \gg k_z$  in Figure \ref{cp:a}. 
This is as one might expect since the disk has a non-zero radial temperature gradient that is coupled with the radial wavenumber. 
The modes representing radial epicyclic overstable oscillations of Section \ref{subsec_ep} show a preference for the $k_z \gg k_R$ region of wavenumber space. This aspect of the modes is in accord with the results of the local analysis by \citet{lyra14}. However, the two modes exhibit contrasting behaviors in the region $k_R H \simeq k_z H \leq 10$ of wavenumber space. 
The first epicyclic mode as represented by Figure \ref{cp:c} appears to be strongest for the above specified portion of wavenumber space. 
In contrast, the second epicyclic mode represented by Figure \ref{cp:b} is completely quenched in the said region. 
Note, however, that the inclusion of vertical structure in the computations quickly damp both these modes as pointed out in Section \ref{subsec_vso}. 
A global or semi-global analysis can potentially shed more light on the relevance of the radial epicyclic modes. 
In Figure \ref{cp:d}, we find that the vertical overstable mode dominates the $k_R \gg k_z$ region. 
A predisposition for that part of wavenumber space is also in accordance with expectation since destabilization by vertical shear 
occurs most effectively through radially narrow, vertically stretched perturbations \citep{ngu13,cmmp14, bl15,ly15}.

\section{Potential Relevance to Protoplanetary Disks}
\label{sec_PPD}

All the unstable modes that we find in the previous section 
have purely hydrodynamic manifestations and should therefore 
directly apply to protoplanetary disks. The applicability
of the computations including a substantially strong toroidal magnetic field is
not immediately obvious and
requires some elaboration. Below we attempt to argue that 
our results including a magnetic field could be useful in addressing the question of axisymmetric stability
within certain regimes of a disk.

While considering a purely toroidal magnetic field may seem a
simplistic assumption, this provides important advantages that enable
us to make connections to protoplanetary disks, where non-ideal
effects play an important role:
{\it i)} A purely toroidal field allows us to consider well-defined
global equilibria for the disk. Both the global model, and its
stability properties, can reach the hydrodynamic limit continuously.
{\it ii)} The absence of a poloidal field eliminates the standard MRI,
which nevertheless is expected to be significantly quenched in the
bulk of the protoplanetary disk.
{\it iii)} The fact that we can consider strong magnetic fields allows
us to make connections with recent studies which suggest that fields
with these characteristics can develop in certain intermediate regions within 
the disk, where ionization levels are very low.

The general consensus is that a sizeable extent of  a protoplanetary disk
constitutes an environment inhospitable to perfect coupling of
magnetic fields to the disk gas. Yet, the standard accretion rates of
$10^{-7} - 10^{-9} M_{\odot}/$yr associated with such disks are most
easily explained if one resorts to disk turbulence. The MRI is the
most favored candidate by far, when it comes to explaining turbulent
transport in disks. However, efficient operation of the MRI requires levels of ionization
 that are believed to be unattainable in significant
portions of the disk. Several local and global studies have been
undertaken to investigate MHD turbulence in the presence of 
some combination of the Ohmic, Hall and ambipolar effects.
At intermediate densities, the Hall effect is believed to be the
dominant non-ideal effect in a protoplanetary disk
\citep{war99}.

Stratified shearing box simulations including all three non-ideal effects of
ohmic diffusion, ambipolar diffusion and the Hall effect
\citep{lkf14, slka15} have lately challenged the magnetically inactive
protoplanetary disk model. \citet{lkf14} have shown that if the
initial magnetic field and angular velocity vector are aligned, the
Hall effect, specifically the Hall-shear instability, leads to the generation of a strong azimuthal field that
adds to pressure support. They speculate that this field geometry 
could possibly be found within the inner reaches of the disk, 
spanning between 1 and 10 AU. Local non-ideal MHD simulations by
\citet{xnb14a} also arrive at a similar picture. These studies find that
the resulting field configurations generate laminar stresses that can
account for the observed accretion rates, thereby, mitigating the need to
appeal to hydrodynamic turbulence or even disk outflows. 
In a similar departure from a purely hydrostatic disk structure, 
\citet{tbdh14} have appealed to magnetically supported atmospheres, with fields strong
enough to alter the disk scale height, in order to explain the
near-infrared excess seen emanating from disks around Herbig Ae/Be stars. 

Diffusive non-ideal effects eventually smoothens out any gradients in the 
magnetic field distribution. Exact equilibrium disk models supporting non-uniform
field configurations inclusive of non-ideal effects cannot therefore be derived. 
Unlike ohmic and ambipolar effects, the 
Hall effect is not diffusive but tends to rearrange field distributions. 
What is not yet clear, however, is whether the 
strong non-uniform field structures observed in simulations with Hall effect \citep{lkf14, xnb14a, slka15}
can sustain for extended periods. Under such circumstances, equilibrium disk models
such as the one described in Section \ref{sec_basiceqns} may constitute
useful, if not exact, representation of at least limited regions within the disk. The suitability
of employing a \emph{localized} version of such a disk model in a Hall dominated regime 
is aided by the fortuitous circumstance under which the linearized Hall 
term vanishes for axisymmetric perturbations. That is, 

\begin{align}
\label{hall1}
\nabla \times \eta_{\rm H}((\nabla \times \boldsymbol{B}) \times \boldsymbol{B} ) = 0 \, ,
\end{align}
where 
\begin{align}
\eta_{\rm H} = \frac{c}{4 \pi n_{\rm e} e}
\end{align}
is the Hall coefficient taken to be uniform and $c$, $n_{\rm e}$, $e$, is the speed of light, the electron density, 
and the charge of the electron, respectively.
Note also that $\boldsymbol{B} = (B_0 + \delta b_{\phi})\boldsymbol{\hat{\phi}}$ represents the sum of 
the background and perturbed magnetic field in Equation (\ref{hall1}).
Therefore, even though our study does not explicitly account for non-ideal
effects, our results may still shed light into the dynamics of disk
regions where the Hall term dominates.  While
arguably rudimentary, this constitutes a first step towards an analytical
understanding of the stability of disk environments suggested by
the recent non-ideal MHD simulations.

\section{Summary and Discussion}
\label{sec_discussion}

We have conducted a comprehensive axisymmetric local linear mode 
analysis of a stratified, differentially rotating disk threaded by a
toroidal magnetic field in the ideal MHD limit.  We employed a thermal 
relaxation model that allowed us to consider a wide range of physical 
conditions between (and including) the isothermal and adiabatic regimes.
Our approach provides a framework to 
investigate the instabilities that can feed off of radial and vertical 
gradients in the dynamic and thermodynamic structure of the disk
in a systematic way. This enables us to put into context previous
studies that have addressed different aspects of this problem
in isolation.

When thermal relaxation takes infinitely long, we find that a new
criteria determines the local axisymmetric stability of disks in the
presence of a purely toroidal field. For toroidal field configurations
that contributes to pressure support, stronger fields act as a
stabilizing influence. The new stability criterion we have derived
reduces continuously to the Solberg-H{\o}iland criteria in the
hydrodynamic limit.

When thermal relaxation takes place on a finite timescale, 
the most important conclusions of this study can be summarized 
as follows.
The mere presence of a background temperature gradient can 
give rise to overstable acoustic oscillations which dominate
 when thermal relaxation is rapid. 
A radial temperature gradient also gives rise to overstable radial epicyclic or acoustic-inertial
oscillations. Combined with a negative radial entropy gradient, the growth rate 
of this epicyclic mode is amplified for longer cooling times. The negative
radial entropy gradient also leads to the emergence of  
another overstable radial epicyclic mode that
is similarly amplified by buoyancy but is present only
for a narrow range of cooling times. 
The two epicyclic overstabilities have essentially the same eigenmode structure
 but differ with respect to growth rates for shorter cooling times. 
These epicyclic modes have been identified in a slightly different
guise as the convective overstability. We posit that a spurious degeneracy
due to the exclusion of the radial temperature gradient made its
way into previous studies
of the convective overstability. 
As a result, the dependence of the growth rate as a function of cooling time
was not correctly captured in previous analysis. 
We believe that the true nature of these fundamental
modes are best described by the results of our study.  
 Strong vertical shear also leads to the development of 
overstable vertical epicyclic modes. These modes grow at an appreciable
rate and are also present only for rather short cooling times. 
The critical value of the cooling time associated with the 
vertical shear  overstability is longer than the corresponding 
time scale for the acoustic overstability. However, the growth 
rate of the overstable acoustic mode is larger than
that of the mode due to vertical shear. 
In short, locally isothermal perturbations
can generally give rise to unstable acoustic, radial epicyclic and vertical epicyclic 
oscillations. These three overstabilities are also present for finite albeit short values of 
cooling times and they are all quickly suppressed beyond 
a critical cooling times that is unique to each mode.
Our results also indicate that the inclusion of both vertical and
radial structure leads to the suppression of the radial epicyclic modes
when vertical shear rates become substantial.

The thermal relaxation or cooling time has been taken as a free
parameter in our analysis. The precise values or ranges of cooling
time is a sensitive function of position, density, temperature,
ionization degree, dust to gas ratio, magnetization and transport
properties in the disk. Estimating the possible ranges of cooling
times is a challenging enterprise and is beyond the scope of this
work. Nonetheless, some amount of effort has been taken, to varying
degrees of sophistication, in deriving theoretical estimates of
cooling times in a disk in the context of the VSI. Both \citet{ngu13}
and \citet{ly15} have attempted to analytically constrain the cooling
times at different locations in the disk. They arrive at the general
conclusion that the VSI should be active in the intermediate regions
of the disk, generally between 5 AU and 50 AU.  Our analysis reveals
that the  overstable acoustic modes have the strongest growth rates
although they require much shorter cooling times. The
overstable epicyclic modes grow with weaker growth rates although they can be
present for a much wider range of cooling times. A thorough analysis
of the extent and ranges of cooling times in a disk guided by
observational constraints is thus warranted.

A global or semi-global analysis of all three overstabilities would certainly shed
more light into the potential relevance of these modes. The non-linear
evolution of these modes also merits further study.

\acknowledgements
We acknowledge useful discussions with Frank Shu, Jacob Simon, Min-Kai Lin, Richard Nelson, Andrew Youdin, Thomas Berlok, Colin McNally, Tobias Heinemann. The research leading to these results has received funding from the European Research Council under the European Union's Seventh Framework Programme (FP/2007-2013) under ERC grant agreement 306614.  M.E.P. also acknowledges support from the Young Investigator Programme of the Villum Foundation.

\bibliography{bibliography} 

\end{document}